\documentclass[iop,apj,tighten]{emulateapj}
\usepackage[breaklinks,colorlinks, urlcolor=blue,citecolor=blue,linkcolor=blue]{hyperref}
\usepackage{color}
\usepackage{epsfig}
\usepackage{rotating}
\usepackage{amsmath}
\usepackage{footnote}
\usepackage{setspace}
\usepackage{courier}
\usepackage{tabularx}
\usepackage{ctable, dashrule}

\newcommand\tab[1][1cm]{\hspace*{#1}}

\definecolor{amethyst}{rgb}{0.6, 0.4, 0.8}

\newcommand{\beq}{\begin{equation}}
\newcommand{\eeq}{\end{equation}}

\newcommand{\avgSFR}{\overline{\raisebox{0pt}[1.2\height]{SFR}}}
\newcommand{\SFR}{\mathrm{SFR}}

\newcommand{\fqcen}{f_\mathrm{Q}^\mathrm{cen}}
\newcommand{\zinit}{z_\mathrm{initial}}
\newcommand{\taucen}{\tau_\mathrm{Q}^\mathrm{cen}}

\begin{document}

\title{Star Formation Quenching Timescale of Central Galaxies in a Hierarchical Universe} 

\author{ChangHoon~Hahn\altaffilmark{1}, 
Jeremy L.~Tinker\altaffilmark{1}, 
Andrew R.~Wetzel\altaffilmark{2,3,4}}
\altaffiltext{1}{Center for Cosmology and Particle Physics, Department of Physics, New York University, 4 Washington Place, New York, NY 10003; chh327@nyu.edu}
\altaffiltext{2}{TAPIR, California Institute of Technology, Pasadena, CA USA}
\altaffiltext{3}{Carnegie Observatories, Pasadena, CA USA}
\altaffiltext{4}{Department of Physics, University of California, Davis, CA USA}
\begin{abstract}
Central galaxies make up the majority of the galaxy population, 
including the majority of the quiescent population at 
$\mathcal{M}_* > 10^{10}\mathrm{M}_\odot$. Thus, the mechanism(s) 
responsible for quenching central galaxies plays a crucial role 
in galaxy evolution as whole.
We combine a high resolution cosmological $N$-body simulation 
with observed evolutionary trends of the ``star formation main sequence,''
quiescent fraction, and stellar mass function at $z < 1$ to 
construct a model that statistically tracks the star formation 
histories and quenching of central galaxies. 
Comparing this model to the distribution of central galaxy 
star formation rates in a group catalog of the SDSS 
Data Release 7, we constrain the timescales over 
which physical processes cease star formation in central galaxies.
Over the stellar mass range $10^{9.5}$ to $10^{11} \mathrm{M}_\odot$ 
we infer quenching e-folding times that span $1.5$ to $0.5\; \mathrm{Gyr}$ 
with more massive central galaxies quenching faster. 
For $\mathcal{M}_* = 10^{10.5}\mathrm{M}_\odot$, this implies a total
migration time of $\sim 4~\mathrm{Gyrs}$ from the star formation main sequence
to quiescence. Compared to satellites, central galaxies take 
$\sim 2~\mathrm{Gyrs}$ longer to quench their star formation, 
suggesting that different mechanisms are responsible for 
quenching centrals versus satellites. Finally, the central galaxy 
quenching timescale we infer provides key constraints for proposed star formation 
quenching mechanisms. Our timescale is generally consistent 
with gas depletion timescales predicted by quenching through 
strangulation. However, the exact physical mechanism(s) 
responsible for this still remain unclear.
\end{abstract}

\section{Introduction}
Observations of galaxies using large galaxy surveys such as the 
Sloan Digital Sky Survey (SDSS; \citealt{York:2000aa}), 
Cosmic Evolution Survey (COSMOS; \citealt{Scoville:2007aa}), 
and the PRIsm MUlti-object Survey (PRIMUS;  \citealt{Coil:2011aa, Cool:2013aa}) 
have firmly established a global view of galaxy properties out to $z \sim 1$. 
Galaxies are broadly divided into two main classes: star forming and quiescent. 
Star forming galaxies are blue in color, forming stars, and typically 
disk-like in morphology. Meanwhile quiescent galaxies are red in color, 
have little to no star formation, and typically have elliptical morphologies 
(\citealt{Kauffmann:2003aa, Blanton:2003aa, Baldry:2006aa, 
Wyder:2007aa, Moustakas:2013aa}; for a recent review see \citealt{Blanton:2009aa}). 

%Star forming galaxies at $z < 1$ (and even beyond) exhibit a tight  correlation between their stellar mass and star formation rate (SFR).  This is referred to in the literature as the ``star formation main sequence'' (SFMS). The SFMS spans several orders of magnitude in stellar mass, has a constant scatter in  $\log\mathrm{SFR}$ of  $\sim 0.3\;\mathrm{dex}$, and decreases significantly in SFR  since $z \sim 1$ (\citealt{Noeske:2007aa, Elbaz:2007aa,  Daddi:2007aa, Salim:2007aa, Whitaker:2012aa, Lee:2015aa}). In fact,  the decline in star formation of the SFMS is likely responsible  for the observed decline of the global star formation in the Universe  (\citealt{Hopkins:2006aa, Madau:2014aa}). 

Over the period $z < 1$, detailed observations of the stellar 
mass functions (SMF) reveal a significant decline in the number density 
of massive star forming galaxies accompanied by an increase in the 
number density of quiescent galaxies 
(\citealt{Blanton:2006aa, Borch:2006aa, Bundy:2006aa, Moustakas:2013aa}). 
The growth of the quiescent fraction with cosmic time also reflects 
this change in galaxy population 
(\citealt{Peng:2010aa, Tinker:2013aa, Hahn:2015aa}). 
Imprints of galaxy environment on the quiescent fraction 
(\citealt{Hubble:1936aa, Oemler:1974aa, Dressler:1980aa, Hermit:1996aa}; 
for a recent review see \citealt{Blanton:2009aa}) suggest that there 
is a significant correlation between environment and the 
cessation of star formation. In comparison to the field, high density 
environments have a higher quiescent fraction.
However, observations find quiescent galaxies in the field 
(\citealt{Baldry:2006aa,Tinker:2011aa,Geha:2012aa}), 
at least for galaxies with stellar mass down to 
$10^9\mathrm{M}_\odot$ (\citealt{Geha:2012aa}), and as 
\cite{Hahn:2015aa} finds using PRIMUS, the quiescent fraction in 
both high density environments and the field increase significantly 
over time. 

Furthermore, galaxy environment is a subjective and heterogeneously 
defined quantity in the literature (\citealt{Muldrew:2012aa}). 
It can, however, be more objectively determined within 
the halo occupation context, which labels galaxies as `centrals' and `satellites'
%In galaxy formation models within the hierarchical structure formation  framework of a $\Lambda$CDM universe, galaxies are typically identified   in, physically motivated, `central' and `satellite' environments 
(\citealt{Zheng:2005aa, Weinmann:2006aa, Blanton:2007ab, Tinker:2011aa}). 
Central galaxies reside at the core of their host halos while 
satellite galaxies orbit around. %and their host halo have fallen into the central host halo. 
During their infall, satellite galaxies are likely to experience 
environmentally driven mechanisms such as ram pressure stripping 
(\citealt{Gunn:1972aa, Bekki:2009aa}),  strangulation 
(\citealt{Larson:1980aa, Balogh:2000aa}), or 
harassment (\citealt{Moore:1998aa}). 

Central galaxies, within this context, are thought to cease their 
star formation through internal processes -- numerous mechanisms have been 
proposed and demonstrated on semi-analytic models (SAMs) 
and hydrodynamic simulations.
One common proposal explains
that hot gaseous coronae form in halos with masses above 
$\sim 10^{12}\mathrm{M}_\odot$ via virial shocks, which starve galaxies 
of cool gas required to fuel star formation 
(\citealt{Birnboim:2003aa, Keres:2005aa,Croton:2006aa,Cattaneo:2006aa, Dekel:2006aa}). 
Other have proposed galaxy merger induced starbursts and subsequent
supermassive blackhole growth as possible mechanisms
(\citealt{Springel:2005aa, DiMatteo:2005aa, Hopkins:2006aa, Hopkins:2008ab, Hopkins:2008aa}). 
Feedback from accreting 
active galactic nuclei (AGN) has also been suggested to contribute to 
quenching (sometimes in conjunction with other mechanisms;
\citealt{Croton:2006aa,Cattaneo:2006aa,Gabor:2011aa}); so has internal 
morphological instabilities in the galactic disk or bar 
(\citealt{Cole:2000aa, Martig:2009aa}). With so many proposed mechanisms  
available, observational constraints are critical to test them.

%Several works have utilized the observed global trends of galaxy populations in 
%order to construct empirical models for galaxy star formation histories
%and quenching. \cite{Schawinski:2014aa}, for instance, use the galaxy color
%magnitude diagram in order to broadly constrain the star formation histories 
%and quenching timescales of morphologically classified early and late type 
%galaxies. For satellite galaxies, \cite{Wetzel:2013aa} quantify the  
%star formation histories and quenching timescales in a statistical and 
%empirical manner. Then using observed SSFR distribution of satellite 
%galaxies, they constrain the quenching timescale of satellites and 
%convincingly illustrate the success of a ``delay-then-rapid'' 
%quenching model, where a satellite after infall begins to quench 
%rapidly after a significant delay. 

Several works have utilized the observed global trends of galaxy 
populations in order to construct empirical models for galaxy star 
formation histories and quenching (e.g. \citealt{Wetzel:2013aa, 
Schawinski:2014aa, Smethurst:2015aa}).
Central galaxies constitute over $70\%$ of the 
$\mathcal{M}_* > 10^{9.7}\mathrm{M}_\odot$ galaxy population at $z = 0$. 
Moreover, the majority of the quiescent population at  
$\mathcal{M}_* > 10^{10}\mathrm{M}_\odot$ become quiescent as 
centrals (\citealt{Wetzel:2013aa}). The quenching of central 
galaxies plays a critical role in the evolution of massive galaxies. 
In this paper, we take a similar approach as \cite{Wetzel:2013aa} but 
for central galaxies. 
\cite{Wetzel:2013aa}
quantify the star formation histories and quenching timescales in 
a statistical and empirical manner. Then using the observed SSFR 
distribution of satellite galaxies, they constrain the quenching 
timescale of satellites and illustrate the success 
of a ``delay-then-rapid'' quenching model, where a satellite begins
to quench rapidly only after a significant delay time after it 
infalls onto its central halo.

Extending to centrals, we use the global trends of 
the central galaxy population at $z < 1$ in order 
to construct a similarly statistical and empirical model for the star 
formation histories of central galaxies. While the initial conditions 
of the satellite galaxies in \cite{Wetzel:2013aa} (at the times
of their infall) are taken from observed trends of the central 
galaxy population, 
%which \cite{Wetzel:2013aa} treat as an ansatz, 
our model for central galaxies must actually reproduce all of the 
multifaceted observations. This requires us to construct
a more comprehensive model that marries all the significant 
observational trends. Then by comparing the mock 
catalogs generated using our model to observations, we constrain 
the star formation histories and quenching timescales of central galaxies. 
Quantifying the timescales of the physical mechanisms that quench 
star formation, not only gives us a means for discerning the numerous
different proposed mechanisms, but it also provides important insights 
into the overall evolution of galaxies. 

We begin first in \S \ref{sec:sdss} by describing the observed 
central galaxy catalog at $z \approx 0$ that we construct from 
SDSS Data Release 7. Next, we describe the cosmological $N$-body simulation 
used to create a central galaxy mock catalog in \S \ref{sec:treepm}. 
We then develop parameterizations of the observed global trends of the
galaxy population and describe how we incorporate them into the mock catalog
in \S \ref{sec:model}. In \S \ref{sec:resultss}, we describe
how we use our model and the observed central galaxy catalog in order to 
infer the quenching timescale of central galaxies. Finally in \S
\ref{sec:discussion} and \S \ref{sec:summary} we discuss the implications of
our results and summarize them. 

\section{Central Galaxies of SDSS DR7} \label{sec:sdss}
We start by selecting a volume-limited sample of galaxies with  $M_r - 5\log(h) < -18$ from the NYU Value-Added Galaxy Catalog (VAGC; 
\citealt{Blanton:2005aa}) of the Sloan Digital Sky Survey Data Release 7
(\citealt{Abazajian:2009aa}) at redshift $z \approx 0.04$ following the 
sample selection of \cite{Tinker:2011aa}. 
The galaxy stellar masses are estimated using the $\mathtt{kcorrect}$ 
code (\citealt{Blanton:2007aa}) assuming a \cite{Chabrier:2003aa} initial 
mass function (IMF). For measurement of galaxy star-formation, we use the 
specific star formation rate (SSFR) from the current release of \cite{Brinchmann:2004aa}
\footnote{http://www.mpa-garching.mpg.de/SDSS/DR7/}. Generally, SSFRs 
$\gtrsim 10^{-11} \mathrm{yr}^{-1}$ are derived from $\mathrm{H}\alpha$ 
emissions, $10^{-11}\gtrsim$ SSFRs $\gtrsim 10^{-12} \mathrm{yr}^{-1}$ 
are derived from a combination of emission lines, and SSFRs
$\lesssim 10^{-12} \mathrm{yr}^{-1}$ are mainly based on $D_n4000$ 
(see discussion in \citealt{Wetzel:2013aa}). 
The spectroscopically derived SSFRs, which accounts for dust-reddening, 
allow us to make more accurate distinctions between star-forming and 
quiescent galaxies than simple color cuts. We note that SSFRs 
$\lesssim 10^{-12} \mathrm{yr}^{-1}$ should only be considered upper 
limits to the true value (\citealt{Salim:2007aa}).  

Next, we identify the central galaxies using the halo-based group-finding 
algorithm from \cite{Tinker:2011aa}. For a detailed description we refer
readers to \cite{Tinker:2011aa, Wetzel:2012aa, Wetzel:2013aa, Wetzel:2014aa}, 
and \cite{Tinker:2016ab}. 
The most massive galaxy of the group is the `central' galaxy and 
the rest are `satellite' galaxies. In any group finding algorithm 
there are misassignments due to projection effects and redshift 
space distortions. \cite{Campbell:2015aa}, quantify both the purity and 
completeness of centrals identified using this group-finding algorithm
at $\sim 80\%$. More importantly, they find that the algorithm 
can robustly identify red and blue centrals and satellites as a 
function of stellar mass and yield a nearly unbiased central red 
fraction, which is the key statistic relevant to our analysis here. 
%Therefore, using this algorithm and only selecting galaxies classified as centrals, we yield an unbiased sample of red and blue centrals.

\section{Simulated Central Galaxy Catalog} \label{sec:treepm}
If we are to understand how central galaxies and their star formation 
evolve, we require simulations over a wide redshift range that allows 
us to examine and track central galaxies within the heirarchical growth 
of their host halos. To do this robustly, we require a cosmological 
N-body simulation that accounts for the complex dynamical processes 
that govern galaxy host halos. In this paper, we use the dissipationless, 
N-body simulation from \cite{Wetzel:2013aa} generated using the 
\cite{White:2002aa} $\mathtt{TreePM}$ code with flat,
$\Lambda$CDM cosmology: $\Omega_\mathrm{m} = 0.274$, 
$\Omega_\mathrm{b} = 0.0457$, $h = 0.7$, $n = 0.95$, and
$\sigma_8 = 0.8$. $2048^3$ particles are evolved in a $250\;\mathrm{Mpc}/h$ 
box with particle mass of $1.98 \times 10^{8}\mathrm{M}_\odot$ and with a Plummer 
equivalent smoothing of $2.5\;\mathrm{kpc}/h$. The initial conditions of 
the simulation at $z = 150$ are generated using second-order Lagrangian 
Perturbation Theory. We refer readers to \cite{Wetzel:2013aa} and 
\cite{Wetzel:2014aa} for a more detailed description of the simulation. 

From the $\mathtt{TreePM}$ simulation, \cite{Wetzel:2013aa} identify 
`host halos' using the Friends-of-Friends (FoF) 
algorithm of \cite{Davis:1985aa} with linking length $b = 0.168$ times the mean 
inter-particle spacing. This groups the simulation particles bound by an isodensity 
contour of $\sim 100\times$ the mean matter density. Within the identified host halos, 
the simulation identifies `subhalos' as overdensities in phase space 
through a 6-dimensional FoF algorithm (\citealt{White:2010aa}). 
\cite{Wetzel:2013aa} then track the host halos and subhalos
across the simulation outputs in order to build merger trees. 
Next, \cite{Wetzel:2013aa} designate the most massive subhalo 
in a newly-formed host halo at a given 
simulation out as the `central' subhalo. A subhalo remains central until it falls 
into a more massive host halo, at which point it becomes a `satellite' subhalo.  
Each subhalo is also assigned a maximum mass $M_\mathrm{peak}$, the 
maximum host halo mass the subhalo has had in its history. 

Using the \cite{Wetzel:2013aa} simulation, we obtain a galaxy 
catalog from the subhalo catalog by assuming that galaxies 
reside at the centers of the subhalos and through subhalo abundance matching 
(SHAM; \citealt{Vale:2006aa, Conroy:2006aa, Yang:2009aa, Wetzel:2012aa, 
Leja:2013aa, Wetzel:2013aa, Wetzel:2014aa}) to assign them stellar masses. 
SHAM assumes a one-to-one mapping that preserves the rank 
ordering between subhalo $M_\mathrm{peak}$ and stellar mass, $\mathcal{M}_*$ 
of its galaxy: $n(>M_\mathrm{peak}) = n(> \mathcal{M}_*)$. Through SHAM, we can 
assign galaxy stellar masses to subhalos based on observed stellar mass function 
(SMF) at the redshifts of the simulation outputs. Galaxy stellar masses 
independently at each snapshot.
This allows us to not only track the history 
of the subhalo, but also track the evolution of galaxy stellar masses through 
their SHAM stellar masses at each snapshot.

%%%%%%%%%%%%%%%%%%%%%%%%%%%%%%%%%%%%%%%%%%%%%%%%%%%%%%%%%%%%%%%%
% Figure: SHAM SMF evolution 
%%%%%%%%%%%%%%%%%%%%%%%%%%%%%%%%%%%%%%%%%%%%%%%%%%%%%%%%%%%%%%%%
\begin{figure}
\begin{center}
\includegraphics[scale=0.45]{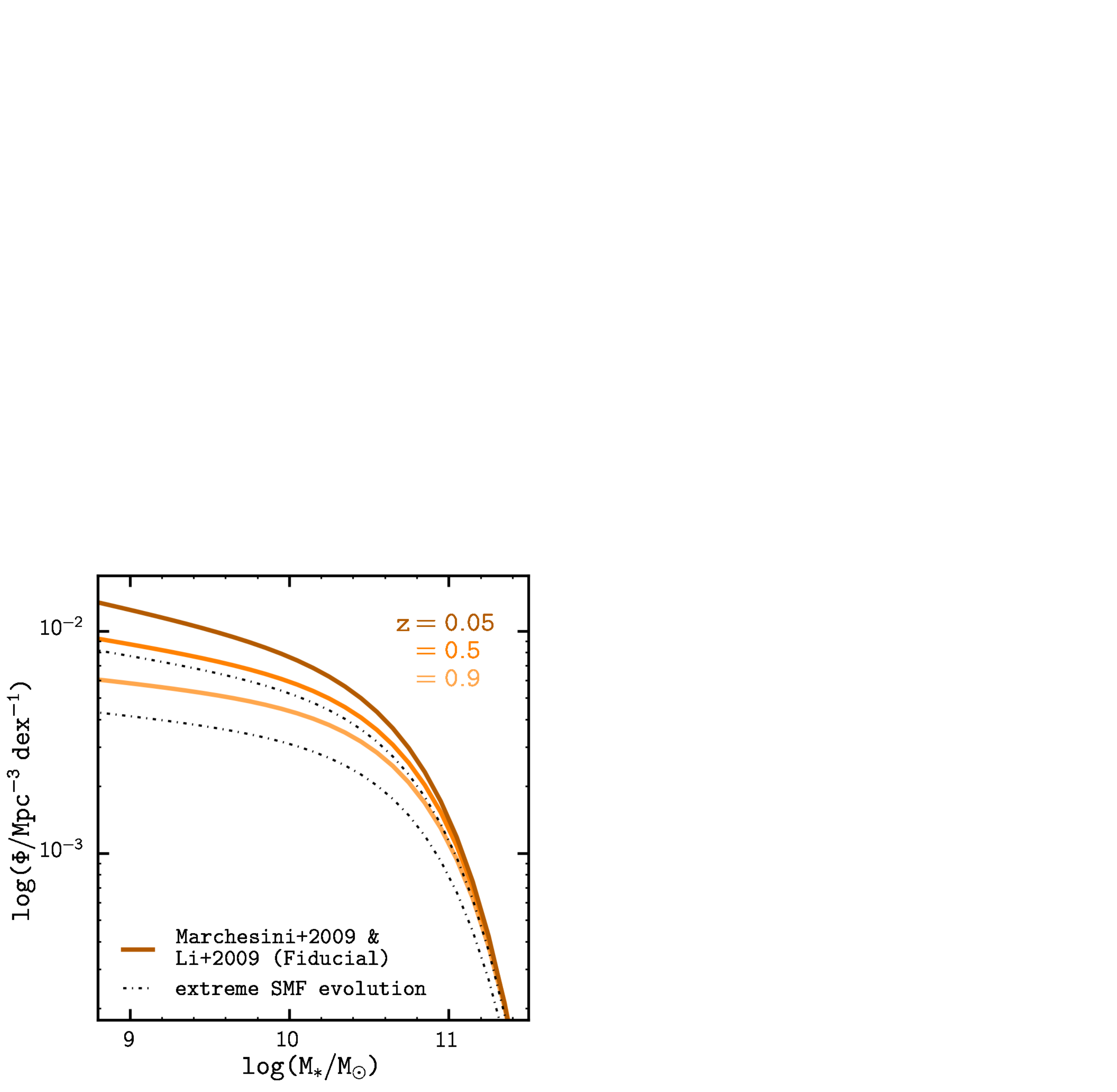}
\caption{The stellar mass function (SMF) that we use in 
our subhalo abundance matching (SHAM) prescription to construct 
galaxy catalogs from the \cite{Wetzel:2013aa} $\mathtt{TreePM}$ 
simulation (\S \ref{sec:treepm}). For our fiducial 
SMF (solid), we use the \cite{Li:2009aa} SMF at $z = 0.05$ and 
interpolate between the \cite{Li:2009aa} SMF and the 
\cite{Marchesini:2009aa} $z = 1.6$ SMF for $z > 0.05$. To illustrate 
the evolution, we plot the SMF at $z=0.05, 0.5$, and $0.9$.
We also plot a SMF parameterization using an ``extreme''
model of SMF evolution (dashed-dotted), in which the amplitude
of the SMF at $z = 1.2$ is half the amplitude of the fiducial SMF. 
We later use this extreme model to ensure that the results in this 
work remain robust over different degrees of SMF evolution at 
$z > 0.05$.}
\label{fig:smf_evol}
\end{center}
\end{figure}

%%%%%%%%%%%%%%%%%%%%%%%%%%%%%%%%%%%%%%%%%%%%%%%%%%%%%%%%%%%%%%%%
% Figure: SHAM SMF evolution 
%%%%%%%%%%%%%%%%%%%%%%%%%%%%%%%%%%%%%%%%%%%%%%%%%%%%%%%%%%%%%%%%
\begin{figure}
\begin{center}
\includegraphics[scale=0.45]{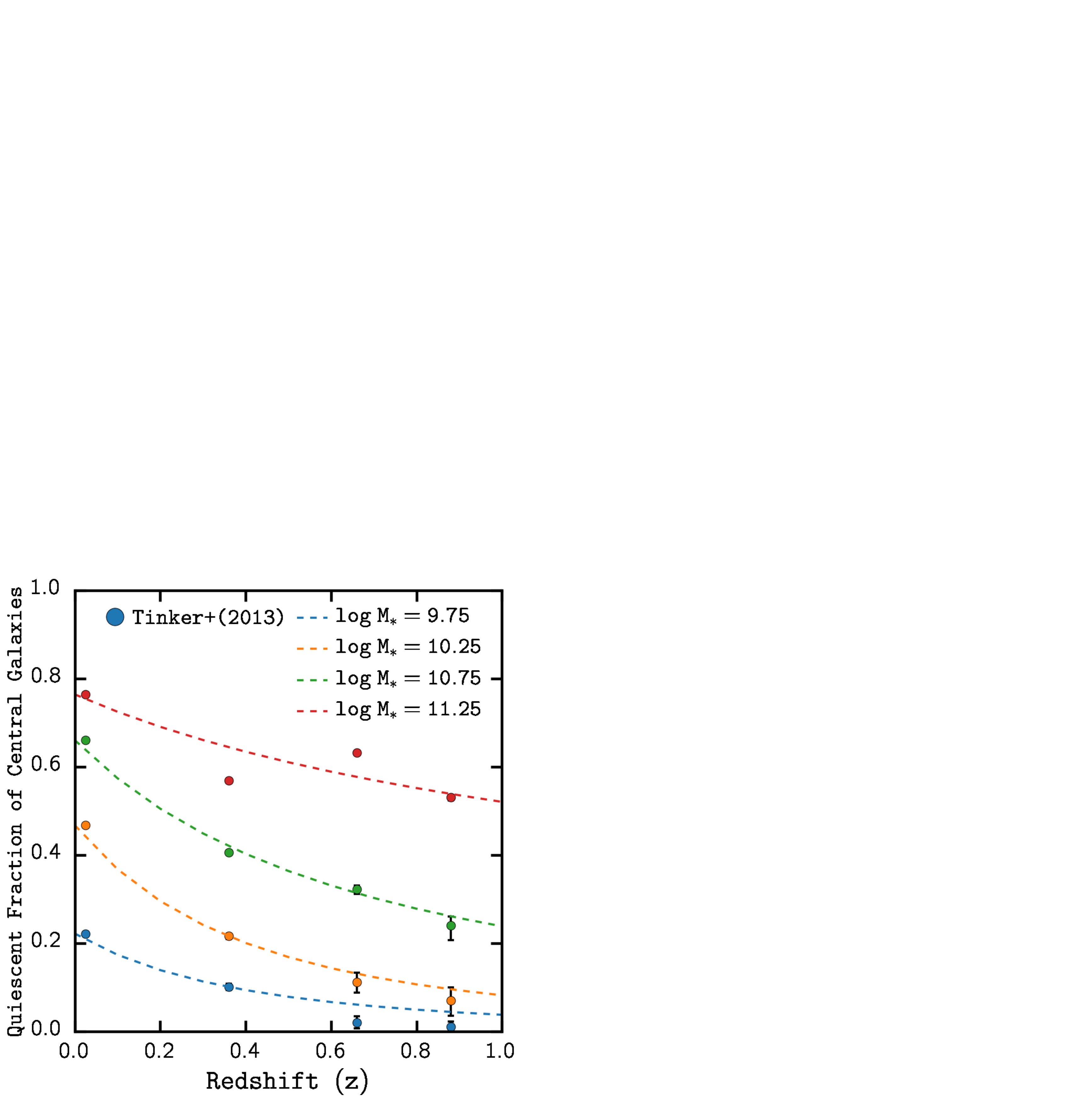}
\caption{The quiescent fraction of central galaxies, $\fqcen$, 
at $z < 1$ in different stellar mass bins. We compare our 
parameterzation of $\fqcen$ (Eq. \ref{fig:fqcen}) using the 
best-fit parameter values listed in Table \ref{tab:fixed_params} (dashed) 
to the $\fqcen$ measurements from \cite{Tinker:2013aa} (scatter). 
For our parameterization, we fit $\fqcen$ at $z = 0.$ using central 
galaxies of the SDSS DR7 group catalog and fit $\alpha(\mathcal{M}_*)$, 
which dictates the redshift dependence, from the redshift evolution  
of the \cite{Tinker:2013aa} measurements.}
\label{fig:fqcen}
\end{center}
\end{figure}
%%%%%%%%%%%%%%%%%%%%%%%%%%%%%%%%%%%%%%%%%%%%%%%%%%%%%%%%%%%%%%%%

For our SHAM prescription, we use the SMF of \cite{Li:2009aa} at the 
lowest redshift $z = 0.05$. \cite{Li:2009aa} is based on the same 
SDSS NYU-VAGC sample as the SDSS DR7 group catalog we describe in 
\S \ref{sec:sdss}. At higher redshifts, we interpolate between the 
\cite{Li:2009aa} SMF and the \cite{Marchesini:2009aa} SMF at $z = 1.6$ 
to obtain the SMF at the simulation output redshifts. This produces 
SMFs that increase significantly and monotonically over $z < 1$ for 
$\mathcal{M}_* < 10^{11} \mathrm{M}_\odot$ but insignificantly for 
$\mathcal{M}_* > 10^{11} \mathrm{M}_\odot$. 
We choose the \cite{Marchesini:2009aa} SMF, amongst others, because 
it produces interpolated SMFs that monotonically increase at $z < 1$. 
At $z \sim 1$, the interpolated SMF we use is consistent (within 
the $1\sigma$ uncertainties) with more recent measurements from 
\cite{Muzzin:2013aa} and \cite{Ilbert:2013aa}. 

In Figure \ref{fig:smf_evol}, 
we illustrate the evolution of the SMFs that we use for our 
SHAM prescription (solid) for $z = 0.05, 0.5$, and $0.9$.
Recently, using PRIMUS, \cite{Moustakas:2013aa} found little 
evolution in the SMF for $z < 1$ at all mass ranges. Although 
previous works such as \cite{Bundy:2006aa} find otherwise. 
To ensure that our results do not depend on our choice of the SMFs, 
later in \S \ref{sec:results}, we repeat our analysis using 
SMFs with no evolution (i.e. \citealt{Li:2009aa} SMF throughout $0< z <1$)
and with ``extreme'' evolution for $z > 0.05$ (dash-dotted in Figure 
\ref{fig:smf_evol}), in which the amplitude of the SMF at $z = 1.2$ 
is approximately half the amplitude of the fiducial SMF at $z = 1.2$.
Furthermore, while the simplest version of SHAM assumes a one-to-one 
correspondence between $M_\mathrm{peak}$ and $\mathcal{M}_*$, observations 
suggest that there is a scatter of $\sim 0.2\;\mathrm{dex}$
in this relation (\citealt{Zheng:2007aa, Yang:2008aa, More:2009aa, Gu:2016aa}).
Hence, we apply a $0.2\;\mathrm{dex}$ log-normal scatter in $\mathcal{M}_*$ 
at fixed $M_\mathrm{peak}$ in our SHAM prescription at each snapshot 
independently. 

So far, we have subhalos populated with galaxies and their 
stellar mass at each of the $15$ simulation outputs spanning 
the redshift $0.05 < z < 1$. 
For our sample, we restrict ourselves to galaxies classified 
as centrals by the simulation. And also to ones that are in 
both the $z\sim 0.05$ and $z\sim 1$ snapshots. This removes 
$< 3\%$ of central galaxies with $\mathcal{M}_* > 10^{9.5}~\mathrm{M}_\odot$ 
in the $z \sim 0.05$ snapshot. Our sample inevitably includes ``back splash'' 
or ``ejected'' satellite galaxies \citep{Wetzel:2014aa}, misclassified 
as centrals. Excluding these galaxies, however, has a negligible 
impact on our results. We also note that while we do not have 
an explicit prescription for stellar mass growth from mergers, 
based on SHAM, the stellar mass growth traces the merger induced 
subhalo growth. As we discuss later in detail, mounting evidence 
disfavor merger driven quenching as the trigger of star formation 
quenching, so our treatment of mergers do not impact our quenching
timescale results. 
In summary, we construct from our simulation a catalog of central 
galaxies whose stellar mass and halo mass is be traced through 
the redshift range $0.05 < z < 1$.

\section{Star Formation in Central Galaxies} \label{sec:model}
The $\mathtt{TreePM}$ simulation (\S \ref{sec:treepm}) provides a framework to examine 
the evolution of central galaxies within the $\Lambda$CDM hierarchical 
structure formation 
of the Universe. In order to determine the quenching timescale of central galaxies, 
we incorporate the evolution of star-formation within this framework so that 
the star formation of the simulated central galaxies reproduce observed trends. 
More specifically, we implement star formation in central galaxies to reproduce
the observed evolution of the quiescent fraction and star-forming main sequence. 

We begin in \S \ref{sec:obvs} by describing our paramertization 
of the observed quiescent fraction and SFMS evolutionary trends at 
$z < 1$. Afterwards, we describe the initial SFR assignment of the
central galaxies in the simulation at the $z = 1$ snapshot in \S \ref{sec:sfr_assign}. 
Then in \S \ref{sec:sfms_evol} we describe how our model evolves 
the SFRs and quenches these central galaxies. 

%%%%%%%%%%%%%%%%%%%%%%%%%%%%%%%%%%%%%%%%%%%%%%%%%%%%%%%%%%%%%%%%
\subsection{Observations} \label{sec:obvs}
With galaxy surveys like the SDSS, COSMOS, and PRIMUS, observations 
have firmly established that for $z < 2$, galactic properties such as 
color and star formation rate (SFR) have a bimodal distribution 
(\citealt{Baldry:2006aa, Cooper:2007aa, Blanton:2009aa, Moustakas:2013aa}).
As mentioned above, the two main components of this distribution are 
{\em quiescent} galaxies with little star formation, which are redder, 
more massive, and reside in denser environments and {\em star forming} 
galaxies, which are bluer, less massive and more often found in the field. 
Since this bimodality is most likely a result of star formation being 
quenched in galaxies, measurements of the quiescent fraction 
$f_\mathrm{Q}$, the fraction of quiescent galaxies in a population, is often 
used to indicate the overall star-forming property of galaxy populations 
(\citealt{Baldry:2006aa, Drory:2009aa, Cooper:2010aa, Iovino:2010aa, Peng:2010aa, 
Geha:2012aa, Kovac:2014aa, Hahn:2015aa}).

For $z < 1$, observations find that the overall quiescent fraction increases as a 
function of stellar mass and with lower redshift (\citealt{Drory:2009aa, Iovino:2010aa, 
Peng:2010aa, Kovac:2014aa, Hahn:2015aa}). In \cite{Wetzel:2013aa}, they quantify 
this mass and redshift dependence of the quiescent fraction through the parameterization, 
$f_\mathrm{Q} (\mathcal{M}_*, z) = A(\mathcal{M}_*) \times (1+z)^{\alpha(\mathcal{M}_*)}$,
with $A(\mathcal{M}_*)$ and $\alpha(\mathcal{M}_*)$ fit from the quiescent fractions of 
the SDSS DR 7 catalog and the COSMOS survey at $z < 1$ (\citealt{Drory:2009aa}), 
respectively. However, the quiescent fraction evolution is not universal over
all environments (\citealt{Hahn:2015aa}). More specifically, \cite{Tinker:2010aa} 
and \cite{Tinker:2013aa} find distinct quiescent fraction evolutions 
for central and satellite galaxies. 

We focus solely on the central galaxy quiescent fraction. We use the same 
parameterization as the overall quiescent fraction parameterization in \cite{Wetzel:2013aa}: 
\beq \label{eq:fq}
f_\mathrm{Q}^\mathrm{cen}(\mathcal{M}_*, z) = f_\mathrm{Q}^\mathrm{cen}(\mathcal{M}_*, z=0) 
\times (1+z)^{\alpha(\mathcal{M}_*)}
\eeq
where 
%We parameterize $f_\mathrm{Q}^\mathrm{cen}(\mathcal{M}_*, z=0)$ as
\beq 
f_\mathrm{Q}^\mathrm{cen}(\mathcal{M}_*, z=0.) = A_0 + A_1\; \log\mathcal{M}_*,
\eeq
is fit to the SDSS DR7 group catalog central galaxies 
(\S \ref{sec:sdss}). $\alpha(\mathcal{M}_*)$, which dictates
the redshift dependence of Eq. \ref{eq:fq}, is fit using the 
redshift dependence of $\fqcen$ measurements from \cite{Tinker:2013aa}, 
derived from observations of the SMF, galaxy clustering, and galaxy-galaxy
lensing within the COSMOS survey,
in bins of width $\Delta \log\mathcal{M}_* = 0.5\;\mathrm{dex}$. 
In Table \ref{tab:fixed_params}, we list the best fit values for 
the parameters  in Eq. \ref{eq:fq} and in Figure \ref{fig:fqcen} 
we compare our parameterization to the \cite{Tinker:2013aa} 
measurements. 

%\todo{To add a figure or not to add a figure...}
%From the $f_\mathrm{Q}$ of all galaxies, \cite{Wetzel:2013aa} then derives 
%the quiescent fraction specifically for centrals by fixing the ratio between the  
%quiescent fraction of centrals versus satellites 
%($f_\mathrm{Q}^\mathrm{cen}/f_\mathrm{Q}^\mathrm{sat}$) throughout $z < 1$. This is 
%motivated by the fact that the satellite galaxy quiescent fraction is always greater
%than the central galaxy quiescent fraction for the same mass 
%(\citealt{Cooper:2007aa, McGee:2011aa, George:2011aa}) and also from observations that 
%find that the quiescent fraction of galaxies in high density environments increases 
%with decrease in redshift (\citealt{Butcher:1984aa,Poggianti:2006aa,McGee:2011aa,Hahn:2015aa}). 
%If $f_\mathrm{Q}^\mathrm{cen}/f_\mathrm{Q}^\mathrm{sat}$ is fixed, then the redshift 
%dependence of $f_\mathrm{Q}^\mathrm{cen}$ is equivalent to the redshift dependence of
%the overall quiescent fraction. So, the central galaxy quiescent fraction can then be 
%parameterized as 

Observations of galaxy populations also find a tight correlation between the SFRs of 
star-forming galaxies and their stellar masses, which is referred 
to in the literature as the ``star formation main sequence'' 
(SFMS; \citealt{Noeske:2007aa, Oliver:2010aa, Karim:2011aa, Moustakas:2013aa}). 
Star-forming galaxies with higher stellar masses have higher SFRs. 
Roughly, this mass dependence can be characterized by a power law, 
$\mathrm{SFR} \propto \mathcal{M}^{\beta}$ and for a given stellar mass, 
SFRs follows a log-normal distribution (\citealt{Noeske:2007aa, Lee:2015aa}). 
Over cosmic time, this tight correlation decreases in SFR but has 
a constant scatter with $\sigma_{\log\mathrm{SFR}} \sim 0.3 \; \mathrm{dex}$ 
(\citealt{Noeske:2007aa, Elbaz:2007aa, Daddi:2007aa, Salim:2007aa, 
Whitaker:2012aa, Lee:2015aa}), In fact this decline SFR of star-forming 
galaxies in the SFMS is likely responsible for the remarkable decline 
of star formation in the Universe (\citealt{Hopkins:2006aa, 
Behroozi:2013aa, Madau:2014aa}).

Following the typical power-law parameterization of the SFMS, we construct a flexible  
parameterization that depends on mass and redshift. For a given stellar mass and 
redshift, the mean SFR of the SFMS is given by 
\beq \label{eq:sfr_ms}
\avgSFR_\mathrm{MS}(\mathcal{M}_*, z) = A_\mathrm{SDSS}
\left(\frac{\mathcal{M}_*}{10^{10.5} \mathrm{M}_\odot} \right)^{\beta_\mathcal{M}}
10^{\beta_z (z - 0.05)}.
\eeq
$A_\mathrm{SDSS}$ is the SFR of the SFMS for the SDSS group catalog at
$\mathcal{M}_* = 10^{10.5} \mathrm{M}_\odot$. We determine 
$\beta_\mathcal{M}$ from fitting Eq. \ref{eq:sfr_ms} to the SFMS of 
the SDSS group catalog ($z = 0.05$). Then we determine $\beta_z$ such 
that the redshift dependence of our estimate of cosmic star formation 
rate,   
\beq \label{eq:rho_sfr}
\rho_\mathrm{SFR}(z) \propto \int \left(1 - f^\mathrm{cen}_\mathrm{Q}\right)
\mathrm{SFR}_\mathrm{MS}(\mathcal{M},z)   
\Phi(\mathcal{M}, z) \; \mathrm{d}\mathcal{M}, 
\eeq
is consistent with the redshift dependence of the cosmic star formation 
rate observations at $z < 1$ (\citealt{Behroozi:2013aa}). 
$\Phi(\mathcal{M},z)$ and $f_\mathrm{Q}^\mathrm{cen}$ in 
Eq.~\ref{eq:rho_sfr}, are the SMF used in the SHAM procedure 
(\S \ref{sec:treepm}) and the central galaxy quiescent fraction 
(Eq. \ref{eq:fq}). 
This agreement in redshift dependence ensures the observational 
consistency between the SMF and the cosmic star formation density 
evolution, which \cite{Behroozi:2013aa} find. 
We list the best fit values to 
$A_\mathrm{SDSS}, \beta_\mathcal{M}$, and $\beta_z$ in 
Table~\ref{tab:fixed_params}. Our $\beta_\mathcal{M}$ and 
$\beta_z$ values are consistent with similar parameterizations in the literature
(\citealt{Salim:2007aa, Moustakas:2013aa, Lee:2015aa}). 

%Using the parameterizations of the central galaxy quiescent fraction and the evolution 
%of the SFMS, in the rest of the section we describe our model designed to track 
%the star formation histories of central galaxies and constrain their quenching 
%timescales. 
%%%%%%%%%%%%%%%%%%%%%%%%%%%%%%%%%%%%%%%%%%%%%%%%%%%%%%%%%%%%%%%%
% Figure: SFR assignment at z=z_i  
%%%%%%%%%%%%%%%%%%%%%%%%%%%%%%%%%%%%%%%%%%%%%%%%%%%%%%%%%%%%%%%%
%\begin{figure}
%\begin{center}
%\includegraphics[scale=0.5]{assignSFR_SSFR.pdf}
%\caption{We present examples of the initial SSFR distribution at 
%$\zinit = 1.08$ of our model described in Section \ref{sec:sfr_assign}. 
%Each distribution uses a different set of initial green valley parameters which gives 
%different green valley fractions: 
%$f_{GV} = 0.2(\log\; \mathcal{M}_* - 10.5)-0.2$ (black solid), 
%$0.4(\log\; \mathcal{M}_* - 10.5)+0.2$ (red dashed), and $0$ (blue dot-dashed). 
%}
%\label{fig:assignSFR}
%\end{center}
%\end{figure}

%%%%%%%%%%%%%%%%%%%%%%%%%%%%%%%%%%%%%%%%%%%%%%%%%%%%%%%%%%%%%%%%
% Figure: SFH demonstration 
%%%%%%%%%%%%%%%%%%%%%%%%%%%%%%%%%%%%%%%%%%%%%%%%%%%%%%%%%%%%%%%%
\begin{figure}
\begin{center}
\includegraphics[scale=0.5]{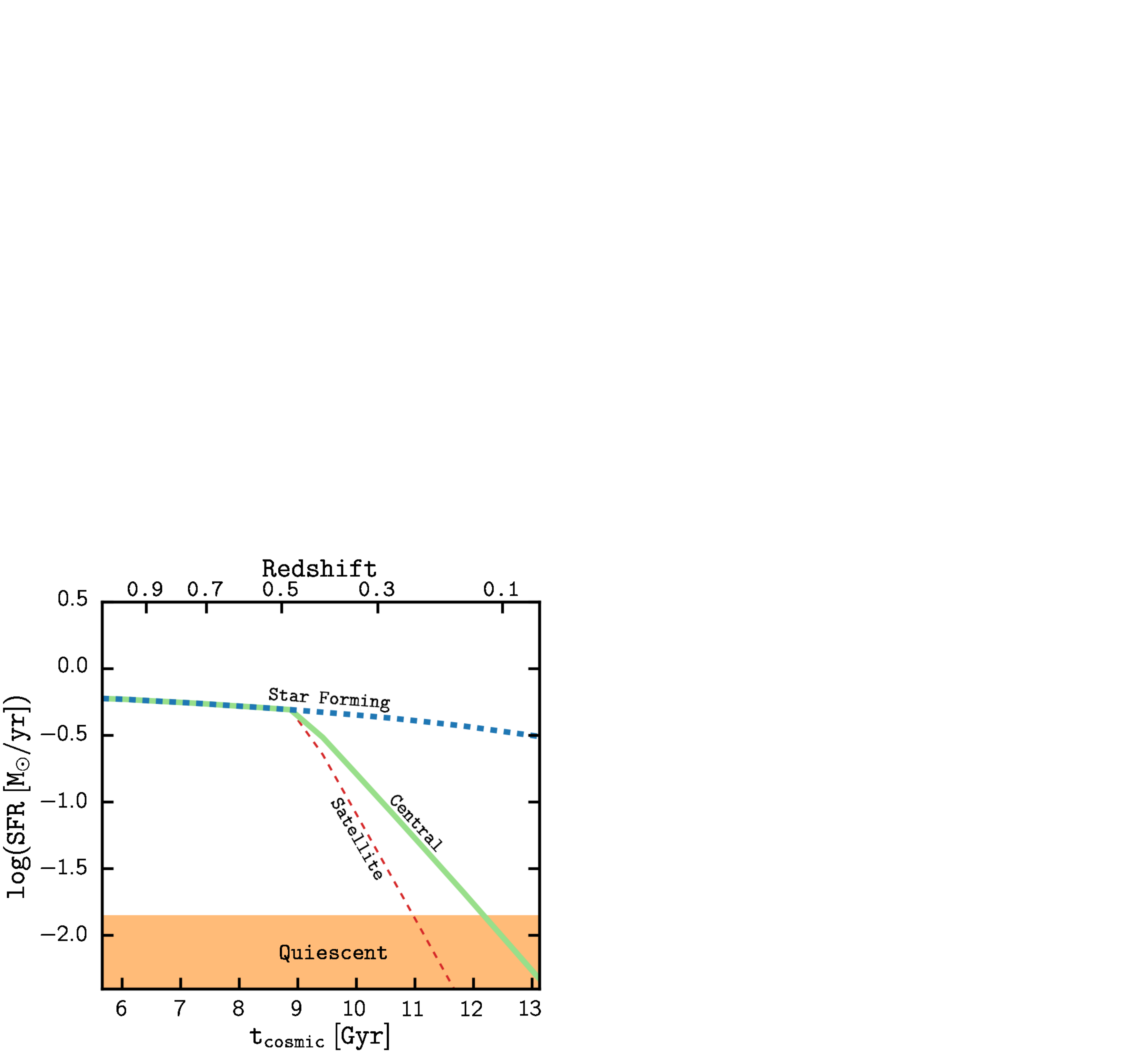}
\caption{Schematic diagram that illustrate the star formation evolution of 
 central galaxies in our model (\S \ref{sec:sfms_evol}). 
We plot SFR as a function of $t_\mathrm{cosmic}$ for a star forming galaxy 
(blue dashed), a star-forming galaxy that quenches at 
$t_\mathrm{Q, start} = 9\;\mathrm{Gyr}$ (green) and mark the general region 
of quiescent galaxies (orange). Central galaxies while they are star-forming 
have SFRs that evolve with the SFMS, which decreases with cosmic 
time. When star-forming central galaxies quench, their SFR decreases 
exponentially with $t_\mathrm{cosmic}$. The quenching timescale, 
$\tau_\mathrm{Q}^\mathrm{cen}$, we constrain in our analysis dictates 
how rapidly these galaxies quench based on Eq. \ref{eq:quenching}. 
For comparison we also plot the SFR evolution of a satellite 
with the same mass using the \cite{Wetzel:2013aa} quenching timescales 
(red dashed).}
\label{fig:SFHdemo}
\end{center}
\end{figure}

%%%%%%%%%%%%%%%%%%%%%%%%%%%%%%%%%%%%%%%%%%%%%%%%%%%%%%%%%%%%%%%%
% Section: Assigning Star Formation Rates
%%%%%%%%%%%%%%%%%%%%%%%%%%%%%%%%%%%%%%%%%%%%%%%%%%%%%%%%%%%%%%%%
\subsection{Assigning Star Formation Rates} \label{sec:sfr_assign}
The first output of the $\mathtt{TreePM}$ simulation that we 
utilize is at $z_\mathrm{initial} = 1.08$. We designate the central
galaxies of this snapshot as quenching, star-forming or quiescent
and assign SFRs to them as the initial conditions of our model. 
The SFR assignment are based on the observed galaxy bimodality, 
the SFMS, and quiescent fraction at $z_\mathrm{initial}$ as we 
detail below.

First, we classify a fraction of the central galaxies as 
``{\em quenching}'' galaxies -- galaxies that reside in the 
green valley, which are in the transitional state of becoming 
quiescent from star-forming. Current observations do not provide 
strong constraints on the fraction of galaxies ``quenching''
at $z \sim 1$. So, we use a flexible and mass dependent prescription  
\begin{equation} \label{eq:f_gv}
f_{GV}(\mathcal{M}_*) = A_{GV} (\log\;\mathcal{M}_* - 10.5) + \delta_{GV}
\end{equation}
and marginalize over the nuisance parameters, $A_{GV}$ and $\delta_{GV}$, 
in our analysis. For these designated quenching central galaxies,
we assign SFRs by uniformly sampling between the average SFR of the 
SFMS (Eq.~\ref{eq:sfr_ms}) at $z_\mathrm{initial}$ and 
$\mathrm{SFR}_\mathrm{Q}(\mathcal{M}_*)$, the 
SFR of the quiescent peak of the SDSS central galaxy SSFR 
distribution, which we later detail in this section.  

Next, we classify the remaining $1 - f_\mathrm{GV}$ of the galaxy population  
as either star-forming or quiescent to match $\fqcen(z=\zinit)$. 
Galaxies classified as star-forming, are assigned SFRs based on the 
log normal SFR distribution of the SFMS at $\zinit$ with scatter 
$\sigma_{\log \SFR} \sim 0.3$ 
(\S \ref{sec:obvs}):
\beq
\log \mathrm{SFR}_\mathrm{SF}^{init} = \mathcal{N} (\log \avgSFR_\mathrm{MS}
\left(\mathcal{M}_*, \zinit), \;0.3 \right).
\eeq
where $\mathcal{N}$ represents a Gaussian.
Galaxies classified as quiescent are assigned SFRs based on a 
log-normal distribution centered about $\avgSFR_{\mathrm{Q},init}$ 
with scatter $\sigma_{\log \SFR}^\mathrm{Q}$: 
\beq
\log \mathrm{SFR}_\mathrm{Q}^{init} = 
\mathcal{N}(\avgSFR_{\mathrm{Q}, init}, \sigma_{\log \mathrm{SFR}}^\mathrm{Q})
\eeq
Both $\avgSFR_{\mathrm{Q},init}$ and 
$\sigma_{\log \mathrm{SFR}}^\mathrm{Q}$ are determined 
empirically from the quiescent peak SSFR in the SDSS 
central galaxy SSFR distribution: 
$\avgSFR_{\mathrm{Q}, init} = 0.4\;(\log\mathcal{M}_* - 10.5) - 1.73$
and $\sigma_{\log \mathrm{SFR}}^\mathrm{Q} = 0.18$.
Our aim is solely to empirically reproduce the quiescent peak because the 
SSFR measurements are largely upper limits, so the peak itself is nonphysical 
(\S \ref{sec:sdss}). 

%In Figure \ref{fig:assignSFR}, we plot the SSFR distribution for  three examples of the initial SFR assignment at $\zinit$ (blue,  black, red). Each SSFR distribution has a unique set of green  valley parameters: $(A_{GV}, \delta_{GV}) = (0.2, -0.2),  (0.4, 0.2)$, and $(0., 0.)$. Next we describe how we evolve star formation  in the central galaxies.

%%%%%%%%%%%%%%%%%%%%%%%%%%%%%%%%%%%%%%%%%%%%%%%%%%%%%%%%%%%%%%%%
% Section: SFR Evolution  
%%%%%%%%%%%%%%%%%%%%%%%%%%%%%%%%%%%%%%%%%%%%%%%%%%%%%%%%%%%%%%%%
\subsection{Star Formation Evolution} \label{sec:sfms_evol}
Starting from the initial SFRs of the central galaxies that we just 
assigned, next, we evolve the SFRs in order to reproduce the observed 
evolution of the quiescent fraction and the SFMS (\S \ref{sec:obvs}). 
The central galaxies in our simulation evolve their SFRs as 
star-forming galaxies, quiescent galaxies and quench their 
star-formation. With the focus of this work 
on the quenching timescale of central galaxies, we first discuss 
how we evolve the SFRs of central galaxies that quench within 
our simulation. Then we discuss how we evolve the 
SFRs of central galaxies while they are star-forming and after they 
have quenched their star formation. 

Once a galaxy begins to quench its star formation, its SFR decreases
and, on the SFR-$\mathcal{M}_*$ relation, it migrates from the SFMS 
to the quenched sequence. We designate the time when a galaxy starts 
to quench as $t_{Q, \mathrm{start}}$ and model its decline in SFR
exponentially with a characteristic e-folding time $\tau_Q^\mathrm{cen}$,
which we refer to as the ``central quenching timescale'': 
\beq \label{eq:quenching}
\mathrm{SFR}_\mathrm{Quenching}(t) = \mathrm{SFR}_\mathrm{SF}(t) \times 
\exp \left(-\frac{t-t_{Q,\mathrm{start}}}{\tau^\mathrm{cen}_Q}\right).
\eeq
$\mathrm{SFR}_\mathrm{SF}$ represents the SFR of a star-forming central 
galaxy, which we define later, and $\tau_Q^\mathrm{cen}$ characterizes 
how long quenching mechanism(s) 
take(s) to cease star-formation in a central galaxy. In order to determine 
whether this timescale depends on the stellar mass of the galaxy, 
we include a mass dependence:
\beq \label{eq:tauq}
\tau_Q^\mathrm{cen}(\mathcal{M}_*) = 
A_\tau \left(\log\;\mathcal{M}_* - 11.1 \right) + \delta_\tau.
\eeq
In addition to the SFR evolution after they begin to quench, 
our model must also quantify when and how many star-forming
centrals quench from $\zinit$. 

For satellite galaxies, the moment they start quenching can be
related to the the moment when their host halo is accreted 
into the central galaxy's host halo via a time delay of 
several Gyrs (\citealt{Wetzel:2013aa}). 
However, for central galaxies, the time when they start to 
quench is likely characterized by more complex and stochastic 
mechanisms such as gas depletion from strangulation (\citealt{Peng:2015aa}), 
hot gas quenching (\citealt{Gabor:2010aa,Gabor:2012aa,Gabor:2015aa}) or 
the onset of AGN activity. Fortunately, using the evolution of the quiescent fraction, 
we can statistically model the number of star-forming centrals 
that quenches throughout the simulation. We use a Monte Carlo 
prescription that utilizes a ``quenching probability" ($P_Q$) 
to determine which star-forming centrals to quench and when to 
quench them. 
We define $P_Q$, for a stellar mass $\mathcal{M}_*$ at simulation snapshot $i$ as  
\beq \label{eq:pq}
P_Q(\mathcal{M}_*, t_i) = 
\frac{N_Q(\mathcal{M}_*, t_{i+1}) - N_Q(\mathcal{M}_*, t_i)}{N_{SF}(\mathcal{M}_*, t_i)}
\eeq
where $N_Q(\mathcal{M}_*, t)$ and $N_{SF}(\mathcal{M}_*, t)$ 
represent the number of quiescent and star-forming galaxies of 
stellar mass $\mathcal{M}_*$ in the population at $t$. 

In the fiducial case where quenching happens instantaneously and 
the time evolution of the stellar mass function is negligible, the 
quenching probability is given directly by the derivative of the 
quiescent fraction over time: 
\beq \label{eq:pq_fid}
P^{fid}_{Q}(\mathcal{M}, t_i) = \frac{t_{i+1} - t_i}{1 - f_\mathrm{Q}(\mathcal{M}, t_i)} \frac{\mathrm{d}f_\mathrm{Q}}{\mathrm{d}t}.
\eeq
However, to account for the SMF evolution, we introduce a correction 
to Eq.~\ref{eq:pq_fid}: 
\beq
\Delta P_{Q}(\mathcal{M}, t_i) = \frac{N_\mathrm{tot}(\mathcal{M}, t_{i+1})-N_\mathrm{tot}(\mathcal{M}, t_{i})}{N_{SF}(\mathcal{M}, t_i)} f_Q(\mathcal{M}, t_{i+1}).
\eeq
Furthermore, star-forming galaxies do not quench instantaneously. 
This implies that some galaxies can begin their quenching 
but still have high enough SFRs to be misclassified as star-forming
causing a discrepancy between when star-forming galaxies start quenching 
to when they become classified as quiescent. This discrepancy depends 
on the SFRs of the quenching galaxies and the timescales of the quenching 
mechanism. Our ultimate goal is to characterize this timescale and its 
dependence on stellar mass, so any strong assumptions may bias our results. 
Therefore, we include a flexible mass dependent factor parameterized by 
$A_{P_Q}$ and $\delta_{P_Q}$ to the quenching probability prescription:
\beq \label{eq:fpq}
f_{P_Q}(\mathcal{M}) = A_{P_Q} (\log\mathcal{M} - 10.5) + \delta_{P_Q}. 
\eeq
By including this term to the quenching probability, we treat $A_{P_Q}$ 
and $\delta_{P_Q}$ as nuisance parameters, which mitigate any biases. 
Combined, the quenching probability we use is
\beq \label{eq:our_pq}
P_{Q,i}(\mathcal{M}) = f_{P_Q} \left(P^{fid}_{Q,i} + \Delta P_{Q,i} \right).
\eeq
Later in \S \ref{sec:results} we discuss the potential impact of our 
quenching probability parameterization on our results. In practice, at 
each simulation output snapshot $t_i$, a number of star-forming central
galaxies are selected to start quenching based on their assigned quenching
probabilities. We note that our quenching probability prescription quenches 
star-forming galaxies anywhere on the SFMS. 

For quenching galaxies before they start quenching and for 
star-forming galaxies that remain star-forming throughout, 
their star formation histories are dictated by the evolution 
of the SFMS. Therefore, we model the star formation evolution of star-forming 
central galaxies to statistically trace the redshift and mass 
dependence of the SFMS. Recall that the stellar masses of the central galaxies
evolve independently from their star formation histories.
Through our SHAM prescription, the stellar mass growth traces 
the mass accretion of its host subhalo (\S~\ref{sec:treepm}).
Then, for a star-forming central with initial stellar mass 
$\mathcal{M}_0$ at $\zinit$ that evolves to $\mathcal{M}$ 
at $z$ to remain on the SFMS, based on Eq.~(\ref{eq:sfr_ms}), 
the SFR at $\zinit$ must evolve by the following factor  
\beq \label{eq:delta_sfr_ms}
f_\mathrm{MS} = \left(\frac{\mathcal{M}}{\mathcal{M}_0}\right)^{\beta_\mathcal{M}} \times
10^{\beta_z (z - z_0)}
\eeq
where $\beta_\mathcal{M}$ and $\beta_z$ are the fixed parameters 
that characterize the mass and redshift dependence of the SFMS 
(\S~\ref{sec:obvs} and Table~\ref{tab:fixed_params}). So
while central galaxies with $\SFR_0$ at $\zinit$ remain 
star-forming they have, 
\beq
\SFR_\mathrm{SF} = f_\mathrm{MS} \times \SFR_0.
\eeq
This way, star-forming galaxies follow the observed redshift 
evolution and mass dependence of the SFMS. Furthermore, since 
our prescription keeps the relative positions in SFR from 
$\avgSFR_\mathrm{MS}$ constant, it preserve the SFR scatter 
of the SFMS -- matching observations. 

Of course, in reality, the SFHs of star-forming central 
galaxies do not strictly follow a simple parameterization of
the SFMS evolution. The stellar mass growth of
the star-forming centrals is not only related to the growth 
of its host subhalo, as our SHAM prescription assumes, but 
also linked to their SFHs. Observations, however, suggest 
a non-trivial connection between stellar mass growth, SFH, 
and host subhalo growth. For instance, if we estimate the 
stellar masses of star forming galaxy by integrating SFRs over time, 
then the stellar mass growth of star-forming galaxies with the same 
initial stellar mass but different SFR on the SFMS, would diverge 
over time and the final stellar masses will be significantly different. 
In that case, it would be difficult to preserve the SFR scatter 
in SFMS along with its log-normal characteristic. 
Alternatively, if independent of subhalo growth, the SFH linked 
stellar mass growth would cause the stellar mass growth for fixed
halo mass to diverge. This would violate the observed 
scatter in the Stellar Mass to Halo Mass (SMHM) relation 
\citep{Leauthaud:2012aa, Tinker:2013aa, Zu:2015aa, Gu:2016aa}.
Clearly a mechanism such as a ``{\em star formation duty cycle}'' 
is required to consolidate observations of the SMHM and the SFMS. 
For the scope of this paper, however, we find that our above
prescription of statistically evolving the SFRs of 
star-forming galaxies is sufficient and incorporating 
stellar mass growth through integrated SFR with a stochastic  
star forming duty cycle, does not significant impact the 
constraints on the quenching timescales. We will investigate 
star formation duty cycle in star-forming central galaxies 
and the link between stellar mass growth, host halo growth
and SFH in Hahn et al. in prep.

Lastly, central galaxies that are quiescent at $\zinit$ 
or become quiescent during the simulation remain quiescent.
Their SFR evolution is determined only to empirically 
reproduce the quiescent peak of the SSFR distribution at $z=0.05$, similar to the 
initial SFR assignment in \S\ref{sec:sfr_assign}. For 
galaxies that are quiescent at $\zinit$, we evolve the 
SFRs in order to conserve the SSFRs throughout the 
simulation: $\mathrm{SSFR}_\mathrm{Q} = \mathrm{SSFR}_0$, 
the initial SSFR at $\zinit$. Then, 
\beq \label{eq:sfr_q}
\mathrm{SFR}_\mathrm{Q} = \mathrm{SSFR}_0 \times \mathcal{M}_*
\eeq
where $\mathcal{M}_*$ is stellar mass at the simulation outputs 
derived from SHAM. 
For galaxies that become quiescent through 
quenching during the simulation, based on Eq.~\ref{eq:quenching}
their SFRs can decrease significantly enough that their SSFRs 
falls below the lower bound of the SDSS DR7 SSFR measurements.
Since we later compare the SSFR distribution of our model to the 
SSFR distribution of the SDSS DR7 central galaxy catalog, 
the quenching galaxies whose SFRs fall below the SDSS 
lower bound will bias the comparison. Therefore, we impose a 
final quenched SSFR assigned based the quiescent peak of 
the observed SSFR distribution for each galaxy when it 
begins to quench. The SFR of quenching galaxies only
decreases until this final quenched SSFR. Afterwards, the
SFR is evolved to conserve the SSFR (Eq.~\ref{eq:sfr_q}). 

Figure \ref{fig:SFHdemo} qualitatively illustrates the SFR evolution of 
star-forming (blue dashed), quenching (green solid), and quiescent (orange) 
central galaxies as a function of cosmic time throughout the simulation. 
The SFR of star-forming galaxies reflects the SFR evolution of the SFMS 
which decreases with time. The quenching galaxy starts to quench at
at $t_\mathrm{Q, start} = 9\;\mathrm{Gyr}$. Its departure from the SFMS 
is clearly illustrated at $t_\mathrm{cosmic} > 9\;\mathrm{Gyr}$. The slope
of its SFR decline is dictated by the quenching timescale. Since the 
lower bound of the SSFR in the SDSS group catalog does not have any 
physical significance, we broadly mark the region with 
$\log\;\mathrm{SSFR} < \log\;\mathrm{SSFR}_\mathrm{Q} + 
\sigma^\mathrm{Q}_{\log\;\mathrm{SFR}}$ as quiescent in Figure \ref{fig:SFHdemo}. 

\begin{figure}
\begin{center}
\includegraphics[width=0.45\textwidth]{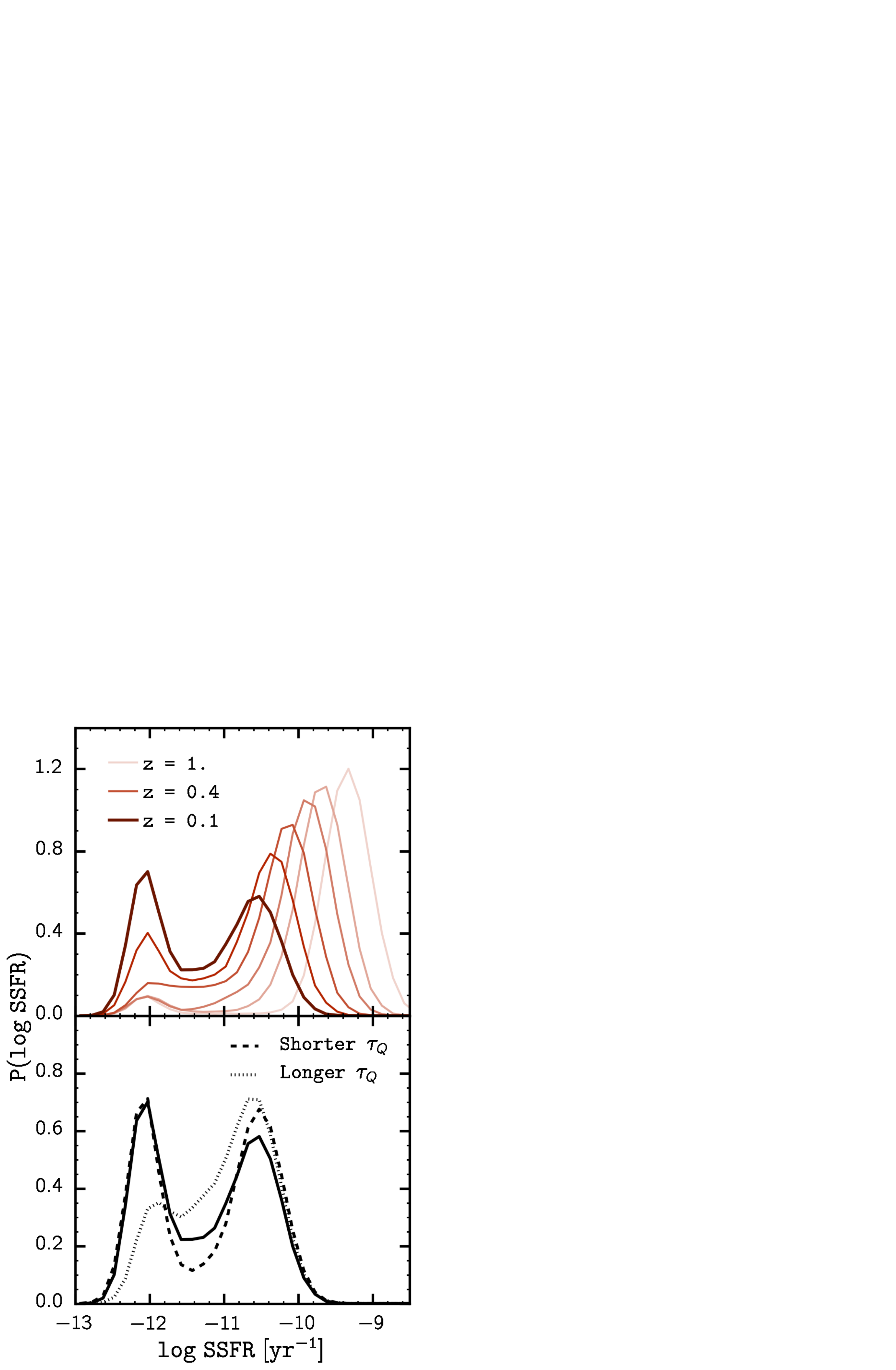}
\caption{
{\bf Top}: The evolution of the SSFR distribution in 
our model (\S \ref{sec:sfr_assign}) for a reasonable set of 
parameter values. The model evolves the SFR of central
galaxies from $z \sim 1$ (light) to $0.05$ (dark) while reproducing 
the observed SFMS and quiescent fraction evolutions. The shift in 
the star forming peak of the SSFR distribution from $z = 1$, 
reflects the overall decline in SFR of the SFMS over time. 
The quiescent fraction evolution is reflected in the growth 
of the quiescent peak accompanied by the decline of the 
star-forming peak. 
{\bf Bottom}: Comparison of the SSFR distribution at $z = 0.05$ 
using a relatively shorter (dashed) and longer (dotted) quenching 
timescale than the above panel (solid). The quenching timescale 
(parameterized by $A_\tau$ and $\delta_\tau$), dictates how long 
quenching central galaxies spend in between the peaks. This is 
ultimately reflected in the height of the green valley. For longer 
quenching timescales, the height of the SSFR distribution green 
valley will higher. For shorter quenching timescales, it will lower.
}
\label{fig:SSFRevol}
\end{center}
\end{figure}

More quantitatively, in the top panel of Figure~\ref{fig:SSFRevol} 
we present the evolution of the SSFR distribution in our model 
(for a reasonable set of model parameter values) to illustrate how 
we track the star formation of central galaxies from $z \sim 1$. 
For this particular set of model parameter values, $f_{GV} \sim 0$ 
within the stellar mass bin.
We plot the SSFR distribution for central galaxies in the stellar 
mass range $[10^{10.1}\mathrm{M}_\odot, 10^{10.5}\mathrm{M}_\odot]$ 
for a number of simulation output snapshots in the redshift range 
$0 < z < 1$ (top; darker with time). 
It demonstrates how our model reproduces the observed evolution 
of the SFMS and quiescent fraction.
With time, the star-forming peak of the SSFR distribution 
decreases in SSFR tracing the SFMS evolution. The amplitude of 
the star-forming peak also decreases and is accompanied by the 
growth of the quiescent peak, reflecting the quiescent 
fraction evolution and the lower bound of SSFR measurements we impose.

In the bottom panel, we compare the SSFR distribution of our model
at $z = 0.05$ using a relatively shorter (dashed) and longer (dotted)
quenching timescale than in the top panel (solid). The quenching
timescale (parameterized by $A_\tau$ and $\delta_\tau$ in 
Eq. \ref{eq:tauq}) dictates how long quenching central galaxies
take to migrate from the SFMS to quiescence. The comparison illustrates 
that the length of the quenching timescale is reflected in the 
``height'' of the SSFR distribution green valley. Longer quenching 
timescales, result in a higher green valley. Shorter quenching timescales, 
result in a lower one.

%%%%%%%%%%%%%%%%%%%%%%%%%%%%%%%%%%%%%%%%%%%%%%%%%%%%%%%%%%%%%%%%
% Table of the parameterizations with fixed parameters
%%%%%%%%%%%%%%%%%%%%%%%%%%%%%%%%%%%%%%%%%%%%%%%%%%%%%%%%%%%%%%%%
\begin{table}
\caption{Parameterizations in the Central Galaxy SFH Model with Fixed Parameters}
\begin{center}
\begin{tabular}{cc} \toprule 
\multicolumn{2}{c}{} \\[-7pt]
\multicolumn{1}{c}{Parameter} & \multicolumn{1}{c}{Value}\\
\hline
\multicolumn{2}{c}{} \\[-7pt]
\multicolumn{2}{c}{Central Galaxy Quiescent Fraction (Eq. \ref{eq:fq})} \\
\multicolumn{2}{c}{$f_\mathrm{Q}^\mathrm{cen}(\mathcal{M}_*, z) = f_\mathrm{Q}^\mathrm{cen}(\mathcal{M}_*, z=0) \times (1+z)^{\alpha(\mathcal{M}_*)}$} \\
\multicolumn{2}{c}{$\tab\tab\;\, = (A_0 + A_1 \log \mathcal{M}_*) \times (1+z)^{\alpha(\mathcal{M}_*)}$} \\[-5pt]
\multicolumn{2}{c}{\hdashrule{8cm}{0.5pt}{3pt 2pt}} \\
\multicolumn{1}{c}{$A_0$} & \multicolumn{1}{c}{$-6.04$}\\
\multicolumn{1}{c}{$A_1$} & \multicolumn{1}{c}{$0.64$}\\ [5pt]
\multicolumn{1}{c}{} & \multicolumn{1}{l}{
$-2.57 \quad \mathcal{M}_* \in [10^{9.5}-10^{10}\mathrm{M}_\odot]$
}\\
\multicolumn{1}{c}{} & \multicolumn{1}{l}{
$-2.52 \quad \mathcal{M}_* \in [10^{10}-10^{10.5}\mathrm{M}_\odot]$
}\\
\multicolumn{1}{c}{$\alpha(\mathcal{M}_*)$} & 
\multicolumn{1}{l}{
$-1.47 \quad \mathcal{M}_* \in [10^{10.5}-10^{11}\mathrm{M}_\odot]$
}\\
\multicolumn{1}{c}{} & \multicolumn{1}{l}{
$-0.55 \quad \mathcal{M}_* \in [10^{11}-10^{11.5}\mathrm{M}_\odot]$
}\\
\multicolumn{1}{c}{} & \multicolumn{1}{l}{
$-0.12 \quad \mathcal{M}_* \in [10^{11.5}-10^{12}\mathrm{M}_\odot]$
}\\
\multicolumn{2}{c}{} \\[-7pt]
\hline
\multicolumn{2}{c}{} \\[-7pt]
\multicolumn{2}{c}{SFMS SFR $z$ and $\mathcal{M}_*$ Dependence (Eq. \ref{eq:sfr_ms})} \\
\multicolumn{2}{c}{$\avgSFR_\mathrm{MS} = A_\mathrm{SDSS} 
\left(\frac{\mathcal{M}_*}{10^{10.5} \mathrm{M}_\odot} 
\right)^{\beta_\mathcal{M}} 10^{\beta_z (z - 0.05)}$} \\[-5pt]
\multicolumn{2}{c}{\hdashrule{8cm}{0.5pt}{3pt 2pt}} \\[3pt]
\multicolumn{1}{c}{$A_\mathrm{SDSS}$} & \multicolumn{1}{c}{$10^{-0.11}\;\mathrm{M}_\odot/\mathrm{yr}$}\\
\multicolumn{1}{c}{$\beta_\mathcal{M}$} & \multicolumn{1}{c}{$0.53$}\\
\multicolumn{1}{c}{$\beta_z$} & \multicolumn{1}{c}{$1.1$}\\[2pt]
\hline
\end{tabular} \label{tab:fixed_params}
\end{center}
We list the parameters and their best-fit values for the
central galaxy quiescent fraction (Eq. \ref{eq:fq}) and SFMS  
SFR redshift and stellar mass dependence (Eq. \ref{eq:sfr_ms}) 
parameterizations. 
$\fqcen(z=0)$ is fit using the central galaxies of the SDSS DR7 
group catalog and the redshift dependence is fit using 
$\fqcen$ measurements from \cite{Tinker:2013aa}. 
Similarly, $\overline{\mathrm{SFR}}_\mathrm{MS}(z=0.05)$ is 
fit using the group catalog while the redshift dependence paramerization 
is fit to reproduce the redshift dependence of the \cite{Behroozi:2013aa} 
cosmic star formation at $z < 1$. 
\bigskip
\end{table}

%%%%%%%%%%%%%%%%%%%%%%%%%%%%%%%%%%%%%%%%%%%%%%%%%%%%%%%%%%%%%%%%
% Figure: SSFR distribution evolution  
%%%%%%%%%%%%%%%%%%%%%%%%%%%%%%%%%%%%%%%%%%%%%%%%%%%%%%%%%%%%%%%%

%%%%%%%%%%%%%%%%%%%%%%%%%%%%%%%%%%%%%%%%%%%%%%%%%%%%%%%%%%%%%%%%
% Table: Free parameters of our parameterization 
%%%%%%%%%%%%%%%%%%%%%%%%%%%%%%%%%%%%%%%%%%%%%%%%%%%%%%%%%%%%%%%%
\begin{table*}
\caption{Parameterizations in the Central Galaxy SFH Model with Free Parameters}
\begin{center}
\begin{tabular}{ccccc} \toprule
\multicolumn{5}{c}{} \\[-7pt]
\multicolumn{1}{c}{Quantity} & \multicolumn{1}{c}{Parameterization} &
\multicolumn{1}{c}{Description} & \multicolumn{1}{c}{Parameter} &
\multicolumn{1}{c}{Prior} \\
\hline
\multicolumn{5}{c}{} \\[-5pt]
% Central Quenching Timescale
\multicolumn{1}{c}{$\tau_Q^\mathrm{cen}(\mathcal{M}_*)$} & 
\multicolumn{1}{c}{$A_\tau (\log\;\mathcal{M}_* - 11.1) + \delta_\tau$} & 
\multicolumn{1}{c}{Central Quenching Timescale in Gyrs (Eq.~\ref{eq:tauq})} & 
\multicolumn{1}{c}{$A_\tau$} & 
\multicolumn{1}{l}{$[-1.5, 0.5]$} \\
\multicolumn{1}{c}{} & \multicolumn{1}{c}{} & \multicolumn{1}{c}{} & 
\multicolumn{1}{c}{$\delta_\tau$} & \multicolumn{1}{l}{$[0.01, 1.5]$} \\[3pt]
% Initial Green Valley Fraction 
\multicolumn{1}{c}{$f_{GV}(\mathcal{M}_*)$} & 
\multicolumn{1}{c}{$A_{GV} (\log\;\mathcal{M}_* - 10.5) + \delta_{GV}$} & 
\multicolumn{1}{c}{Initial $z \approx 1$ Green Valley Fraction (Eq.~\ref{eq:f_gv})} & 
\multicolumn{1}{c}{$A_{GV}$} & 
\multicolumn{1}{l}{$[0., 1.]$} \\
\multicolumn{1}{c}{} & \multicolumn{1}{c}{} & \multicolumn{1}{c}{} & 
\multicolumn{1}{c}{$\delta_{GV}$} & \multicolumn{1}{l}{$[-0.4, 0.6]$} \\[3pt]
% Quenching Probability Factor 
\multicolumn{1}{c}{$f_{P_Q}(\mathcal{M}_*)$} & 
\multicolumn{1}{c}{$A_{P_Q} (\log\mathcal{M}_* - 10.5) + \delta_{P_Q}$} & 
\multicolumn{1}{c}{Quenching Probability Factor (Eq.~\ref{eq:fpq})} & 
\multicolumn{1}{c}{$A_{P_Q}$} & 
\multicolumn{1}{l}{$[-5., 0.]$} \\
\multicolumn{1}{c}{} & \multicolumn{1}{c}{} & \multicolumn{1}{c}{} & 
\multicolumn{1}{c}{$\delta_\tau$} & \multicolumn{1}{l}{$[0.5, 2.5]$} \\[3pt]
\hline
\end{tabular} \label{tab:free_params}
\end{center}
We list the parameterizations of the quenching timescale (Eq. \ref{eq:tauq}), 
the initial $z \approx 1$ green valley fraction (Eq. \ref{eq:f_gv}), 
and the quenching probability factor (Eq.\ref{eq:fpq}) that we use in our 
model (\S\ref{sec:model}). In our Approximate Bayesian Computation 
parameter inference, we constrain the parameters listed in the four 
column. For the prior probability distributions of
these parameters, we use uniform priors with the ranges listed in the 
last column. We note that while we allow $\delta_{GV} < 0$ due to the mass 
dependence of $f_{GV}$, $f_{GV}$ can only be non-negative in our model.
\bigskip
\end{table*}

%%%%%%%%%%%%%%%%%%%%%%%%%%%%%%%%%%%%%%%%%%%%%%%%%%%%%%%%%%%%%%%%
% Figure: ABC Corner plot 
%%%%%%%%%%%%%%%%%%%%%%%%%%%%%%%%%%%%%%%%%%%%%%%%%%%%%%%%%%%%%%%%
\begin{figure*}
\begin{center}
\includegraphics[scale=0.5]{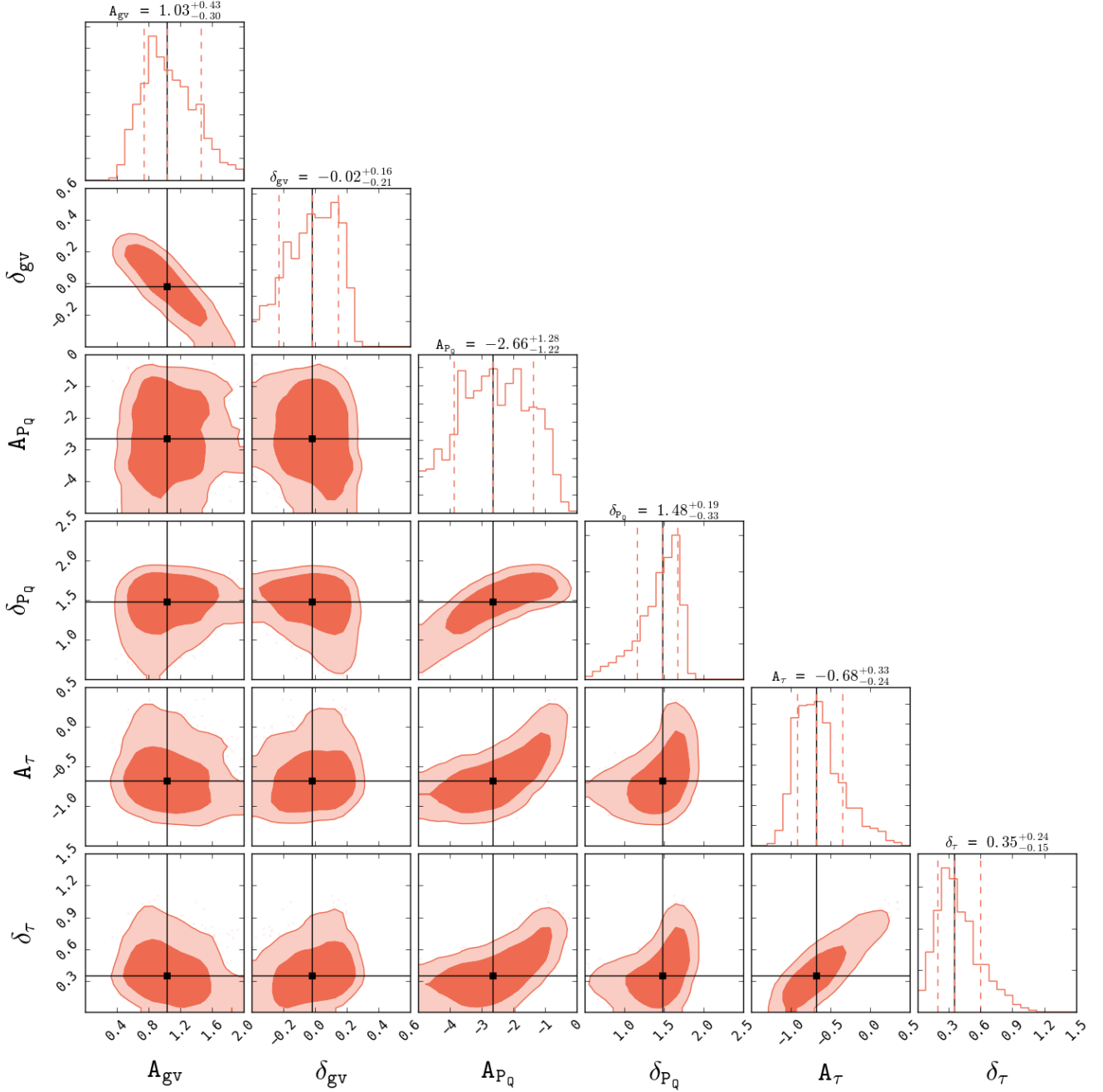}
\caption{We present the constraints we obtain for our model parameters using ABC-PMC. 
The diagonal panels plot posterior distributions of each of our model parameters, while
the off-diagonal panels plot the degeneracies of parameter pairs. For each of the posterior
distributions, we mark the $68\%$ confidence interval (vertical dashed lines). We also 
mark the median of the posterior distributions in all the panels in black.}
\label{fig:abc_post}
\end{center}
\end{figure*}
%%%%%%%%%%%%%%%%%%%%%%%%%%%%%%%%%%%%%%%%%%%%%%%%%%%%%%%%%%%%%%%%
% Figure: ABC SSFR distribution 
%%%%%%%%%%%%%%%%%%%%%%%%%%%%%%%%%%%%%%%%%%%%%%%%%%%%%%%%%%%%%%%%
\begin{figure*}
\begin{center}
\includegraphics[width=\textwidth] {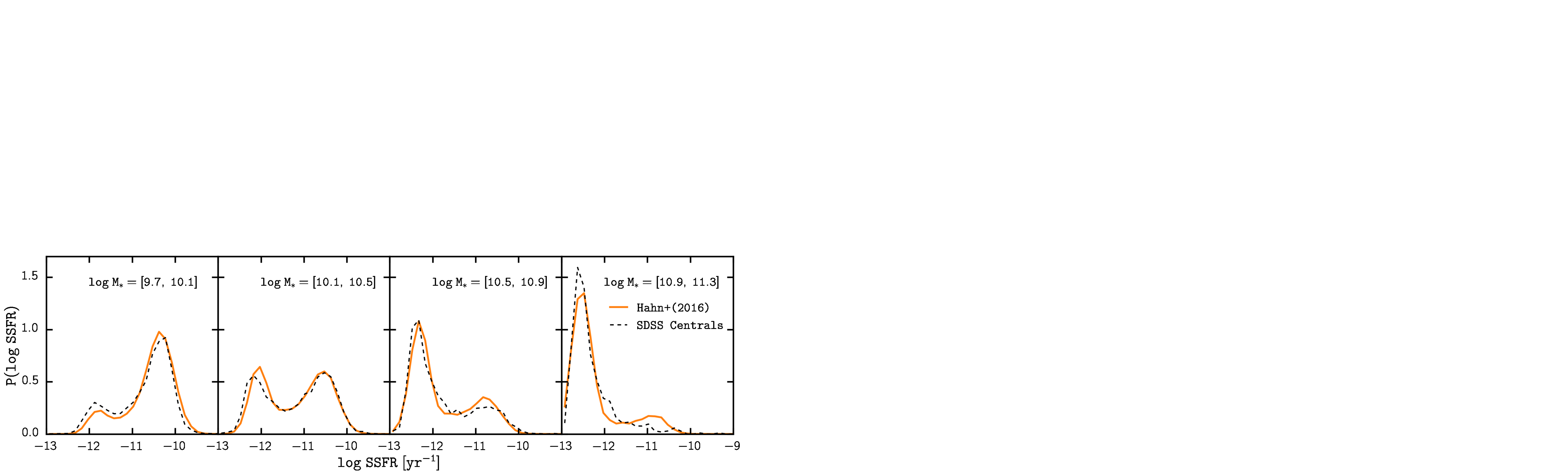}
\caption{Comparison between the SSFR distribution calculated using the 
median of the ABC posterior distribution as the set of model parameters (orange)
and the SSFR distribution of the SDSS DR7 central galaxies (black dash). The SSFR 
distribution from the median of the ABC posterior show good overall agreement. The 
distributions are especially consistent in the transition (green valley) regions, 
which are dictated by the quenching timescale parameters.}
\label{fig:abc_ssfr}
\end{center}
\end{figure*}
%%%%%%%%%%%%%%%%%%%%%%%%%%%%%%%%%%%%%%%%%%%%%%%%%%%%%%%%%%%%%%%%
% Figure: ABC tau distribution 
%%%%%%%%%%%%%%%%%%%%%%%%%%%%%%%%%%%%%%%%%%%%%%%%%%%%%%%%%%%%%%%%
\begin{figure}
\begin{center}
\includegraphics[scale=0.45]{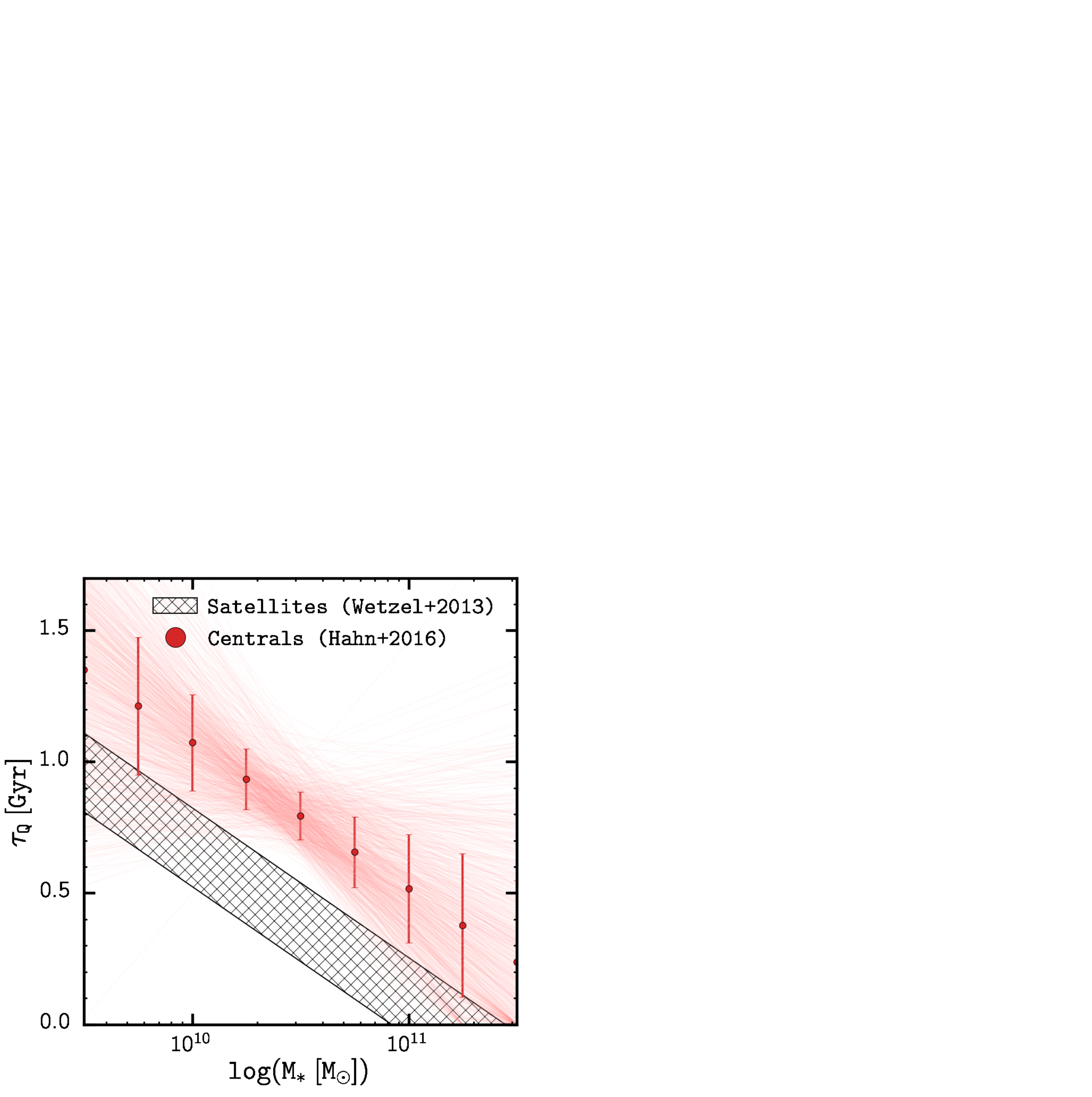}
\caption{Quenching timescale, $\taucen$, of central galaxies (red) 
as a function of stellar mass. We plot $\taucen$ of the median parameter values 
of the ABC posterior distributions (red points) along with $\taucen$ drawn from 
the final iteration ABC parameter pool (faint red lines). For comparison, we also 
plot the satellite quenching timescale of \cite{Wetzel:2014aa} (black dashed). 
%We note that $\taucen$ is a function of the central galaxy stellar mass at  $t_{Q, \mathrm{start}}$ (Eq. \ref{eq:quenching}), while \cite{Wetzel:2013aa}  parameterizes the satellite quenching timescale as a function of stellar mass  at $z = 0.05$. Since the overall stellar mass of the satellite galaxy population grows  with cosmic time, parameterizing the satellite quenching timescale as a function  of stellar mass at $t_{Q, \mathrm{start}}$, would only enhance the difference between the central and satellite quenching timescales. 
The constraints we
get for quenching timescale of central galaxies reveal that central galaxies 
have a significantly longer quenching timescale than satellite galaxies.} 
\label{fig:abc_tau}
\end{center}
\end{figure}

%%%%%%%%%%%%%%%%%%%%%%%%%%%%%%%%%%%%%%%%%%%%%%%%%%%%%%%%%%%%%%%%
% Figure: ABC SSFR tau satellite distribution 
%%%%%%%%%%%%%%%%%%%%%%%%%%%%%%%%%%%%%%%%%%%%%%%%%%%%%%%%%%%%%%%%
\begin{figure*}
\begin{center}
\includegraphics[width=\textwidth]{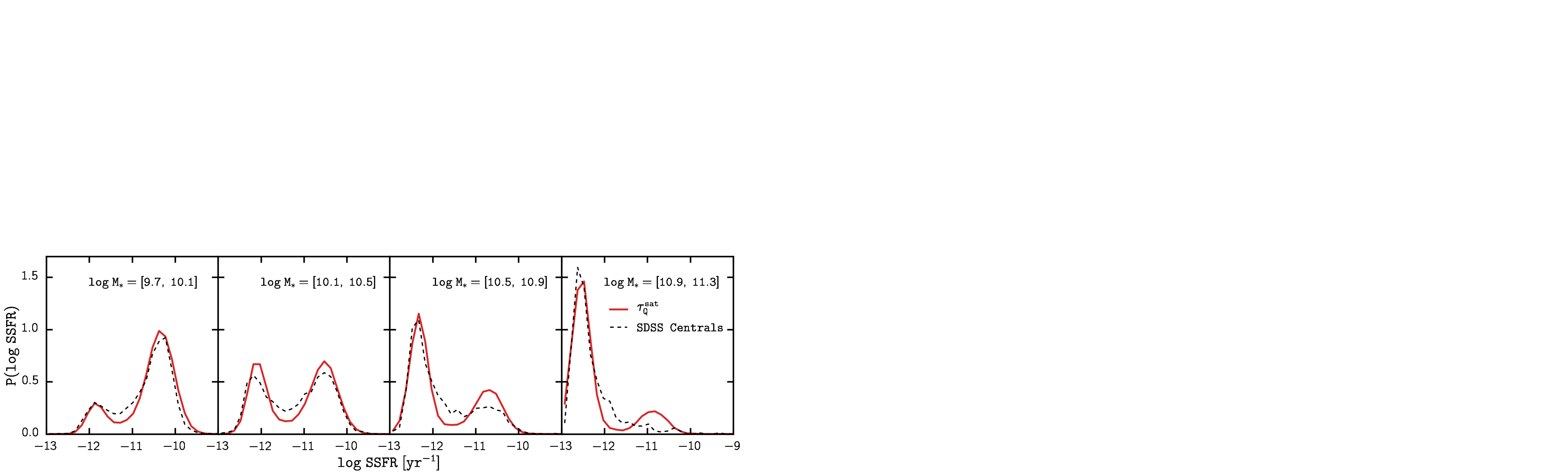}
\caption{The SSFR distribution generated from the median values of the 
parameter constraints obtained from our analysis using \cite{Wetzel:2013aa} 
satellite quenching timescale as the quenching timescale of our 
central galaxies (red) in four stellar mass bins. In each panel, 
we reproduce the quiescent fraction of the SDSS DR7 central galaxies;
however, comparison to the SSFR distribution of the SDSS DR7 centrals 
(black dash) find significant discrepancies in each of the bins. 
The SSFR distribution using satellite quenching timescale have much 
shallower green valley regions as a result of galaxies quenching much 
faster with satellite quenching timescale. {\em This disagreement of 
model predictions for satellites applied to observations of centrals 
clearly demonstrates that centrals require longer quenching timescales 
than satellites.} 
}
\label{fig:ssfr_tau_sat}
\end{center}
\end{figure*}
%Timescales longer than the satellite quenching timescales are necessary to reproduce  the green valley of the SDSS DR7 central galaxy SSFR distribution. 
%%%%%%%%%%%%%%%%%%%%%%%%%%%%%%%%%%%%%%%%%%%%%%%%%%%%%%%%%%%%%%%%
% Figure: tau_Q using different SMF evolution  
\begin{figure}
\begin{center}
\includegraphics[width=0.45\textwidth]{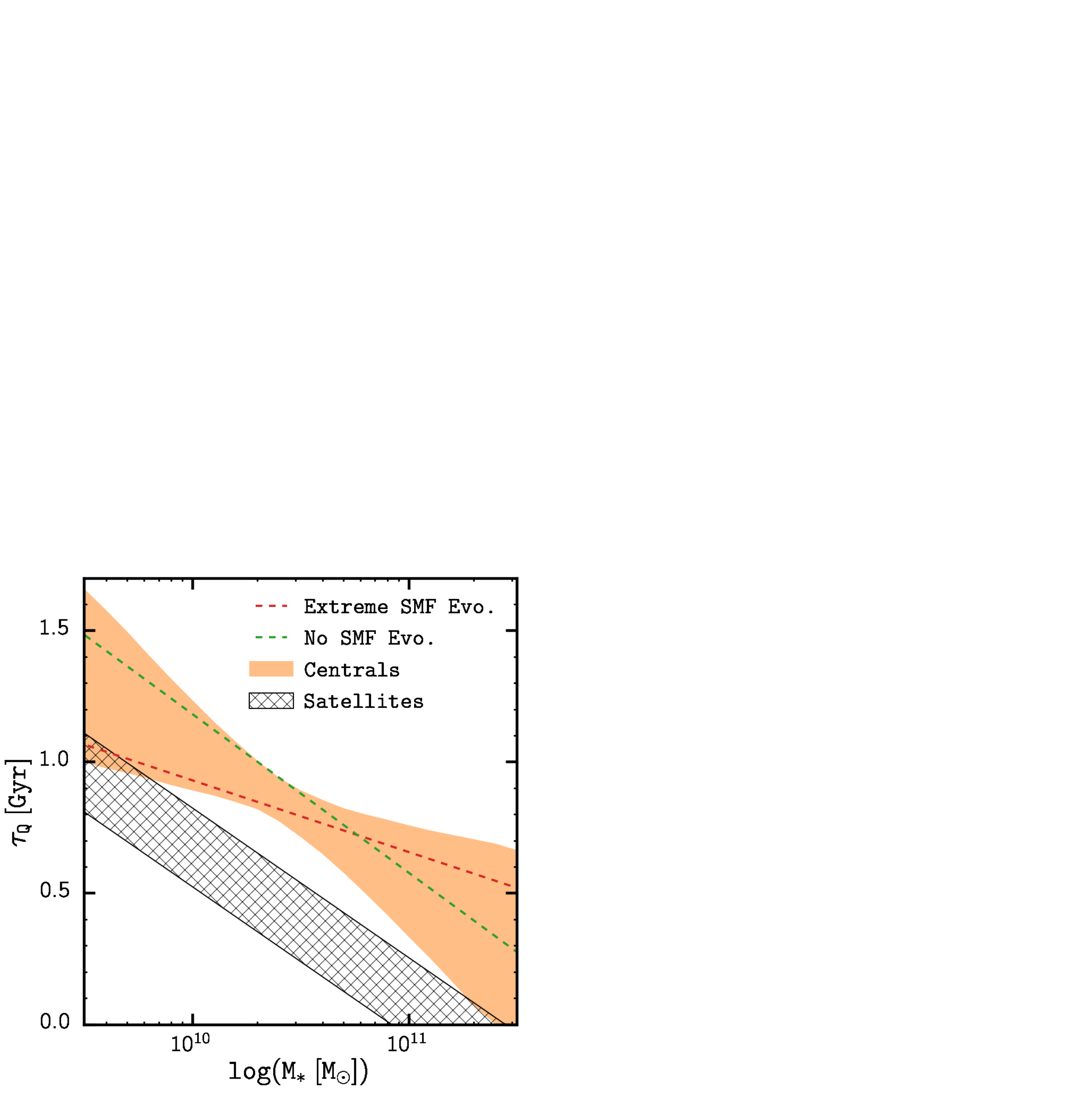}
\caption{Central galaxy quenching timescales ($\taucen$) derived from using 
SMF prescriptions with no SMF evolution (green) and with extreme SMF evolution 
(red) in our analysis. For comparison we include the satellite galaxy quenching 
timescale from \cite{Wetzel:2013aa} and $\taucen$ we obtain using our fiducial 
SMF prescription. Even extreme choices for the SMF evolution is insufficient to 
account for the significant difference between the central and satellite 
quenching timescales. The different SMF evolution mainly impacts the mass 
dependence, not the amplitude of $\taucen$.}
\label{fig:tau_smfevol}
\end{center}
\end{figure}

%%%%%%%%%%%%%%%%%%%%%%%%%%%%%%%%%%%%%%%%%%%%%%%%%%%%%%%%%%%%%%%%
% Central Quenching Timescales  
%%%%%%%%%%%%%%%%%%%%%%%%%%%%%%%%%%%%%%%%%%%%%%%%%%%%%%%%%%%%%%%%
\section{Results} \label{sec:resultss}
Now that we have a model for evolving star formation in central 
galaxies, in this section, we constrain the parameters of the 
model.
%we specifically seek to constrain the  parameters that dictate the quenching timescale. 

\subsection{Approximate Bayesian Computation} \label{sec:abc}
Approximate Bayesian Computation (ABC) is a generative, 
simulation-based inference for robust parameter estimation. 
It has the advantage over standard approaches for parameter 
inference in that it does not require explicit knowledge 
of the likelihood function. It only relies on a simulation 
of observed data and on a metric for the distance between 
the observed data and simulation. It has already been 
effectively used for astronomy and cosmology in the literature 
(\citealt{Cameron:2012aa, Weyant:2013aa, Akeret:2015aa, 
Ishida:2015aa, Lin:2015aa, Lin:2016aa, Hahn:2016aa}, and 
Cisewski et al. in prep.), spanning a wide range of topics.
For our purposes, which is to constrain the quenching timescale 
parameters, we use the observed SSFR distribution and quiescent 
fraction. ABC provides an ideal framework for parameter inference 
without having to specify the explicit likelihood of these 
observables. In practice, we use 
ABC in conjunction with the efficient 
Population Monte Carlo (PMC) importance sampling 
(\citealt{Ishida:2015aa, Hahn:2016aa}). 

ABC requires a number of specific choices for implementation: a 
simulation of the data, a set of prior probability distributions 
for the model parameters, and a distance metric to compare the 
``closeness'' of the simulation to the data. In \S 
\ref{sec:model}, we described our model for the star formation 
evolution of central galaxies. The parameters of our model, which we 
constrain in our ABC analysis are listed in Table \ref{tab:free_params}. 
For the prior probability distributions of the simulation parameters, 
$\{A_\mathrm{GV}, \delta_\mathrm{GV}, A_{P_Q}, \delta_{P_Q}, A_\tau, 
\delta_\tau \}$, we choose uniform priors with conservative ranges also 
listed in Table \ref{tab:free_params}. 
%We note that our prior  distributions require all of the $\delta$ parameters to be  positive. As the  parameterizations in Table \ref{tab:params} illustrate,  This is because  they represent the amplitudes of the green valley fraction,  the quenching probability factor,  and the quenching timescale, which by definition cannot be negative. 

The distance metric in ABC parameter estimation is --- in principle --- a
positive definite function that compares various summary statistics between
between the data and the simulation. It can be a vector with multiple 
components where each component is a distance between one particular 
summary statistic of the data and that of the simulation. For our case, 
the summary statistics we use for our distance metric are the observables
we seek to reproduce with our model: {\em the quiescent fraction evolution and 
SDSS DR7 central galaxy SSFR distribution}. Therefore, we use a two component 
distance metric, $\vec\rho = [\rho_\mathrm{QF}, \rho_\mathrm{SSFR}]$. 

We calculate the first component, $\rho_\mathrm{QF}$, so that our model 
best reproduces the quiescent fraction at multiple snapshots:
\beq \label{eq:rho_qf}
\rho_\mathrm{QF} = \sum\limits_\mathcal{M} \sum\limits_{z' \in \{z\}} 
\left(f^\mathrm{cen}_\mathrm{Q}(\mathcal{M}, z') - f^\mathrm{model}_\mathrm{Q}(\mathcal{M}, z') \right)^2
\eeq
where $\{z\} = \{0.05, 0.16, 0.34$, and $1.08\}$ and $f^\mathrm{cen}_\mathrm{Q}$
is the parameterization of the observed quiescent fraction (Eq. \ref{eq:fq}. 
For $f^\mathrm{model}_\mathrm{Q}$ rather than using the actual evolutionary 
stages of the simulation central galaxies, we measure it using the following 
$\mathrm{SFR} - \mathcal{M}_*$ cut. The $\mathrm{SFR} - \mathcal{M}_*$ cut
is derived from the slope of the SFMS relation (Eq. \ref{eq:sfr_ms}): 
\beq
\log\;\mathrm{SFR}_\mathrm{cut} = \log\;\avgSFR_\mathrm{MS} - 0.9. 
\eeq
If a galaxy SFR is less than $\mathrm{SFR}_\mathrm{cut}$, then it is 
classified as quiescent; otherwise, as star-forming. This classification
is analogous to the quiescent/star-forming classification of 
\cite{Moustakas:2013aa}, which also utilizes the slope of the SFMS. By 
measuring the quiescent fraction of the simulation we are more consistent 
with observations, which have no way of knowing the evolutionary stage
of galaxies beyond their $\mathrm{SFR}$ and $\mathcal{M}_*$. 

Our redshift choices for Eq. \ref{eq:rho_qf} is primarily 
motivated to ensure that our model agrees with the observed 
quiescent fraction throughout the lower redshifts ($z < 0.5$). 
By incorporating the $z' < 0.5$ contributions, we constrain 
$A_\mathrm{P_Q}$ and $\tau_\mathrm{P_Q}$, which dictate the 
quenching probabilities. $\zinit=1.08$ is also included to 
ensure that our initial conditions are consistent with observations. 
%so that our flexible prescription for the green valley  fraction (Eq. \ref{eq:f_gv}) does not compromise the consistency between  $f_\mathrm{Q}^\mathrm{model}$ and $f_\mathrm{Q}^\mathrm{cen}$ at $z_\mathrm{initial}$. 

The second component of our distance metric compares the SSFR distribution 
of the SDSS DR7 central galaxies to that of our model. More specifically, 
\beq \label{eq:rho_ssfr}
\rho_\mathrm{SSFR} = \sum\limits_\mathrm{SSFR} 
\left( P(\mathrm{SSFR})^\mathrm{SDSS} - P(\mathrm{SSFR})^\mathrm{model} \right)^2.
\eeq
As we discuss in \S~\ref{sec:sfms_evol}, the quenching timescale
parameters leave an imprint on the SSFR distribution. So $\rho_\mathrm{SSFR}$ 
successfully serves to constrain $A_\tau$ and $\delta_\tau$. 

Beyond our choice of distance metric, we strictly follow the ABC-PMC 
implementation of \cite{Hahn:2016aa}. For aficionados, we use a median 
distance threshold after each iteration of the PMC and declare 
convergence when the acceptance ratio falls below $1\%$. Once converged, 
the ABC algorithm produces parameter distributions that generate models 
with quiescent fractions and SSFR distributions close to observations. 
Moreover, these parameter distributions predict the posterior distributions 
of the parameters. For further details, we refer readers to \cite{Hahn:2016aa}.

\subsection{Central Galaxy Quenching Timescale} \label{sec:results}
We present the central galaxy quenching timescale constraints 
we obtain using ABC (\S \ref{sec:abc}), in Figure 
\ref{fig:abc_post}. The diagonal panels of the figure plot 
the posterior distribution of each of our model parameters 
with vertical dashed lines marking the median and the $68\%$ 
confidence interval. The off-diagonal panels plot the 
degeneracies between parameter pairs. We also mark the median 
of the posterior distribution for each of the parameters (black). 
The off-diagonal panels illustrate that the initial green valley 
parameters are not degenerate with the other parameters. 
Galaxies that are initially in the green valley quickly 
evolve out of it, so the green valley prescription is mainly 
constrained by the quiescent fraction at $\zinit$. Furthermore, 
the off-diagonal panels that plot the degeneracies between 
the quenching probability parameters and the quenching timescale 
parameters exhibit expected correlation between the parameters: 
the longer the quenching timescale the larger the quenching 
probability correction factor ($f_{P_Q}$). 

We compare the SSFR distribution generated from our model 
using the median model parameter values of the posterior 
distribution (orange) to the SSFR distribution of the SDSS 
DR7 central galaxy catalog (black dashed), in Figure 
\ref{fig:abc_ssfr}. The SSFR distribution are computed for 
four stellar mass bins. We find good agreement between the 
SSFR distributions in each of the bins. More importantly, 
the model with parameters values from the posterior distribution 
is able to successfully reproduce the height of the green valley.

In Figure \ref{fig:abc_tau}, we plot the central galaxy quenching timescale
$\tau^\mathrm{cen}_Q(\mathcal{M})$ corresponding to the median parameter 
values of the posterior (red points) and compare it to the satellite 
quenching timescale in \cite{Wetzel:2013aa}. We also plot 
$\tau^\mathrm{cen}_Q(\mathcal{M})$ for $A_\tau$ and $\delta_\tau$ of the 
final iteration ABC parameter pool (light red lines) and error bars on
median $\tau^\mathrm{cen}_Q$ to represent the 1-sigma values
in stellar mass bins of width $\Delta \log\mathcal{M} = 0.25~\mathrm{dex}$.
The model used in \cite{Wetzel:2013aa} to infer the satellite 
quenching timescale has notable difference from our model. However, 
an analogous analysis reproduces an equivalent satellite quenching timescale.
The comparison of the quenching timescales reveal that both timescales 
exhibit significant mass dependence, which curiously appear to have 
similar slopes. The similarity, however, is difficulty to precisely 
quantify because of the uncertainties in both timescales.
The comparison, above all, illustrates that {\em the quenching timescale 
of central galaxies is significantly longer than the quenching timescale 
of satellites}. 
%Central galaxies on average take significantly longer to  quench their star formation than satellite galaxies. Furthermore, the \cite{Wetzel:2014aa} satellite quenching timescale  is parameterized as function of galaxy stellar mass at $z = 0.05$ while our $\tau^\mathrm{cen}_Q(\mathcal{M})$ (Eq. \ref{eq:quenching}),   is a function of galaxy stellar mass at $t_{Q, \mathrm{start}}$.  Overall, the stellar mass for the satellite population at $t_{Q, \mathrm{start}}$,  is {\em less} massive than the stellar mass at $z = 0.05$. So, qualitatively,  this means that the difference between the quenching timescales is even greater than  what is portrayed in Figure \ref{fig:abc_tau}. 

To determine whether our constraints on the central galaxy 
timescale are robust, we carry out a similar analysis where
we fix the quenching timescale parameters to the satellite quenching 
timescale of \cite{Wetzel:2013aa}. Then we use ABC-PMC with 
Eq. \ref{eq:rho_qf} as the distance metric to constrain the 
parameters $A_\mathrm{GV}$, $\delta_\mathrm{GV}$, $A_{P_Q}$, 
and $\delta_{P_Q}$. In Figure \ref{fig:ssfr_tau_sat}, we plot 
the SSFR distribution generated from median parameter values 
of the parameter constraints and compare it to the SSFR 
distribution of the SDSS DR7 central galaxies. At all stellar 
mass bins, while the quiescent fraction is generally reproduced, 
the height of the green valley for the model using satellite 
quenching timescale is significantly lower than the green valley 
of the SDSS DR7 centrals. 
Therefore, a longer quenching timescale is necessary to 
reproduce the height of green valley for central galaxies.

In our model, we obtain stellar masses of central galaxies 
from the SHAM prescription of host subhalos. As a result, 
the stellar mass evolution of our central galaxies is sensitive 
to the SMF and its evolution. In our SHAM procedure, we formulate 
the SMF based on \cite{Li:2009aa} and \cite{Marchesini:2009aa},
which evolves significantly for 
$\mathcal{M} < 10^{11} \mathrm{M}_\odot$ over $z < 1$. SMF measurements 
from PRIMUS for $z < 1$ in \cite{Moustakas:2013aa}, however, fail to find 
such significant SMF evolution. To confirm whether or not our 
central quenching timescale constraint remains robust over 
different degrees of SMF evolution, we test our results 
with two extreme models of SMF evolution (included in Figure 
\ref{fig:smf_evol}): (1) a model in which the SMF does not 
evolve with time, and (2) a model in which the SMF at $z = 1.2$ 
is roughly half of our fiducial SMF at $z=1.2$. We plot the 
results in Figure \ref{fig:tau_smfevol}.
We plot the $\tau^\mathrm{cen}_Q(\mathcal{M})$ of the median posterior 
parameter values from our analysis using extreme models of SMF evolution. 
While the SMF evolution impacts the mass dependence,  
$\tau^\mathrm{cen}_Q(\mathcal{M})$ remains significantly longer than the 
quenching timescale of satellites.

We also repeat the analysis for different parameterizations of 
$\fqcen$; more specifically, the two $\fqcen$ parameterization 
in \cite{Wetzel:2013aa}. Regardless of the $\fqcen$ 
parameterization, we find that $\tau^\mathrm{cen}_Q(\mathcal{M})$ is 
greater than the satellite quenching timescale. We conclude that 
our central quenching timescale results are robust over the specific 
choices we make in implementing our model.

%%%%%%%%%%%%%%%%%%%%%%%%%%%%%%%%%%%%%%%%%%%%%%%%%%%%%%%%%%%%%%%%
% Discussion 
%%%%%%%%%%%%%%%%%%%%%%%%%%%%%%%%%%%%%%%%%%%%%%%%%%%%%%%%%%%%%%%%
\section{Discussion} \label{sec:discussion}
\subsection{Central versus Satellite Quenching}
One key result of the central galaxy quenching timescales we infer 
is its difference with the satellite galaxy quenching timescale from 
\cite{Wetzel:2013aa}. For the entire stellar masses range probed, 
the quenching timescale of central galaxies is $\sim 0.5\;\mathrm{Gyr}$ 
longer than that of satellite galaxies. This corresponds to central 
galaxies taking approximately $\sim 2 \;\mathrm{Gyrs}$ longer
than satellite galaxies to transition from the SFMS to the quiescent 
peak. Moreover, this difference suggests that {\em quenching 
mechanisms responsible for the cessation of star formation in central 
galaxies are different from the ones in satellite galaxies}. 

\begin{figure}
\begin{center}
\includegraphics[width=0.5\textwidth]{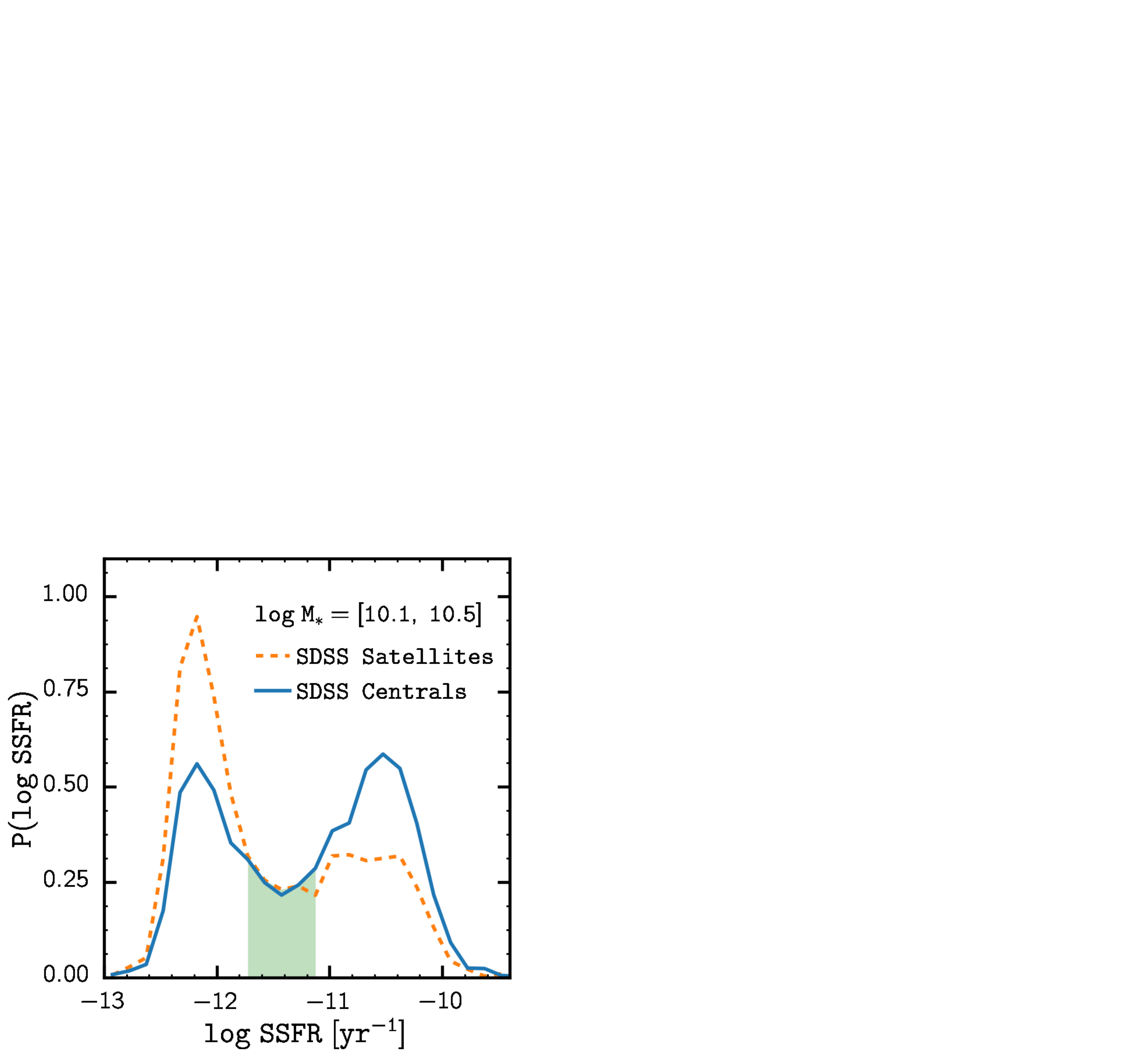}
\caption{SSFR distributions of the central galaxies
versus the satellite galaxies in the SDSS DR7 Group Catalog with
stellar mass between $10^{10.1}$ and $10^{10.5}\mathrm{M}_\odot$. 
Both SSFR distributions have similar green valley heights 
(green shaded region). Since
central galaxies have significantly longer quenching timescales, 
satellite galaxies have a higher rate of transitioning from 
star-forming to quenching than central galaxies.}
\label{fig:sdss_censat_ssfr}
\end{center}
\end{figure}

At a glance, this difference in central and satellite quenching 
timescale is rather unexpected since the SSFR distribution of
central (blue) and satellite (orange) galaxies of the SDSS DR7 
Group Catalog in Figure \ref{fig:sdss_censat_ssfr}, show remarkably 
similar green valley heights.  
However, the similarity in green valley height is not determined 
by the quenching timescale alone. It reflects the combination 
of quenching timescale and the rate that star-forming galaxies 
transition to quenching. Since the satellite quenching timescale
is shorter than that of centrals, star-forming satellites 
transition to quenching at a higher rate than star-forming 
centrals at $z = 0$. The difference in this transition rate is
even higher than what the quenching timescale reflects because 
tidal disruption and mergers preferentially destroy quiescent 
satellite galaxies.  

The implication that satellites and centrals have different quenching 
mechanisms is broadly consistent with the currently favored dichotomy of 
quenching mechanisms: satellite galaxies undergo environmental 
quenching while central galaxies undergo internal quenching. 
It is also consistent with the significant difference in the structural 
properties of quiescent satellites versus centrals \citep{Woo:2016aa}, 
which also suggests different physical pathways for quenching
satellites versus centrals.
Furthermore, it explains the environment dependence in the quiescent fraction 
evolution in recent observations (\citealt{Hahn:2015aa, Darvish:2016aa}). 
Both central and satellite quenching contribute in high density 
environments while only central quenching contributes in the field 
causing the quiescent fraction to increase more significantly in high
density environments. 

Additionally, combined with the \cite{Wetzel:2013aa} 
result that at $\mathcal{M}_* > 10^{10} \mathrm{M}_\odot$ central 
galaxy quenching is the dominant contributor to the growth of the 
quiescent population, we can also characterize mass regimes where 
environmental or internal quenching mechanisms dominate, similar to 
\cite{Peng:2010aa}. 
Below $\mathcal{M}_* < 10^{9} \mathrm{M}_\odot$, satellite quenching 
is the {\em only} mechanism \citep{Geha:2012aa} and internal quenching 
is ineffective. Until $\mathcal{M}_* < 10^{10} \mathrm{M}_\odot$, 
environmental quenching continues to be the dominant mechanism. At 
$\mathcal{M}_* > 10^{10} \mathrm{M}_\odot$ internal quenching 
dominates.

% Figure: P(SSFR) central vs satellite comparison  

\begin{figure}
\begin{center}
\includegraphics[width=0.45\textwidth]{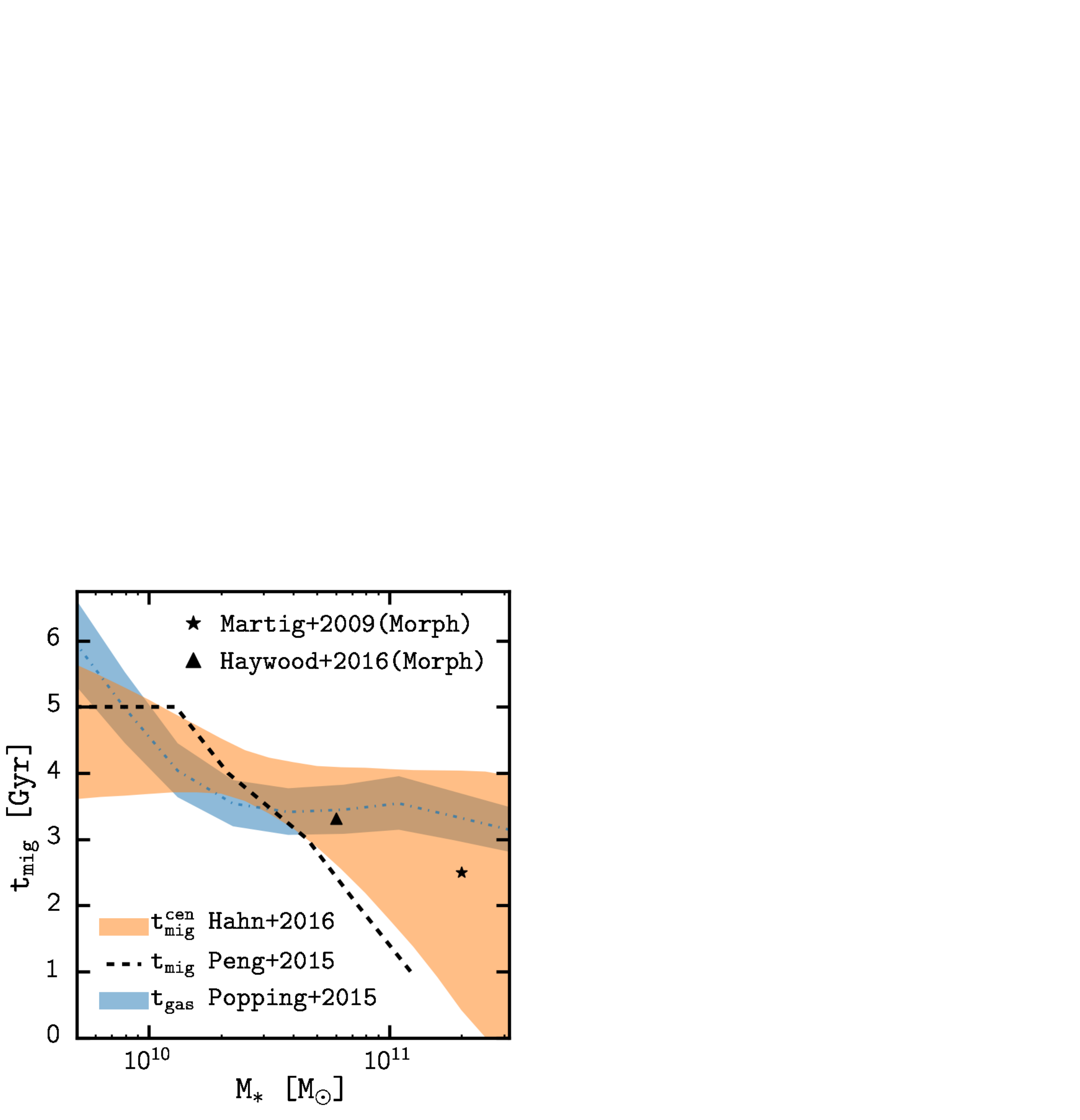}
\caption{
Comparison of the central galaxy quenching migration time 
estimate we infer, ($t_\mathrm{mig}^\mathrm{cen}$; orange) with 
quenching time estimates for gas depletion absent accretion (strangulation)
and morphological quenching. The width represents the 68 \% 
confidence region propagated from the posterior distributions of the 
$\taucen$ parameters. 
For strangulation, we include the gas depletion time at $z = 0.2$ derived 
from the star formation efficiency estimates in \cite{Popping:2015aa} 
(blue dash-dotted). The surrounding blue shaded region plots the range 
of gas depletion times at $z = 0.15$ (longer) to $0.25$ (shorter). 
We also include the quenching migration time inferred from 
the \cite{Peng:2015aa} gas regulation model (dashed). 
For morphological quenching we plot the quenching times taken from 
the star formation histories of the simulated galaxy in 
\cite{Martig:2009aa} (star). We also include the quenching times 
of the Milky Way in \cite{Haywood:2016aa} (triangle). 
The quenching timescale of strangulation exhibit a similar
stellar mass dependence and is generally consistent with our central 
quenching timescales. Although its feasibility 
for a wider galaxy population is unexplored, the quenching timescale
from morphological quenching is in good agreement with our timescale.}
\label{fig:tquench_comp}
\end{center}
\end{figure}

\subsection{Quenching Star Formation in Central Galaxies}  
Numerous physical processes have been proposed in the literature to explain
the quenching of star formation. Observations, however, have yet to identify 
the primary driver of quenching or consistently narrowing down proposed
mechanisms. The quenching timescale we derive for central galaxies provides 
a key constraint for any of the proposed mechanisms. Only processes that agree 
with our central galaxy quenching timescales, can be the main driver for quenching 
star formation in central galaxies. 

Merger driven quenching has often been proposed as a driving mechanism 
of star formation quenching (\citealt{Springel:2005aa, Hopkins:2006ab, 
Hopkins:2008ab, Hopkins:2008aa}). In this proposed mechanism, quenching
is typically driven by gas-rich galaxy mergers which induce starburst and rapid 
black hole growth. 
%\cite{Hopkins:2008ab, Hopkins:2008aa},  for instance, using a model derived from theoretical halo and subhalo mass  functions and empirical halo occupation models along with the ansatz that  star formation is quenched after a gas-rich major merger, argued that merger driven quenching {\em can} explain the growth of the red sequence. When we examine the merger histories of the central galaxies host halos  in our simulation, however, we find that mergers (both minor and major) are  insufficient to account for the central quiescent fraction we observe at  $z = 0.05$. In Figure \ref{fig:merger}, we compare the fraction of central galaxies in our simulation at $z = 0.05$ that have experienced a major  (\todo{some color}) and  minor (\todo{some color}) merger to the  central quiescent fraction of the SDSS DR7 Group Catalog. Major mergers are  defined as a merger between two galaxies with mass ratios less than or equal to  $3:1$ whereas minor mergers are mergers with mass ratios less than or equal to  $10:1$ but greater than $3:1$. For $\mathcal{M} < 10^{11} \mathrm{M}_\odot$,  minor and major mergers, even combined, is insufficient in accounting for  the quiescent fraction. \todo{Update this one figure is made!}
Cosmological hydrodynamics simulations that examine mergers, however, 
conclude that quenching from mergers alone cannot 
produce a realistic red sequence (\citealt{Gabor:2010aa, Gabor:2011aa}.
\cite{Gabor:2011aa} used an on-the-fly 
prescription to identify mergers and halos in order to test
different prescriptions for quenching star formation. In addition to failing 
to produce a realistic red sequence, they find that mergers cannot 
sustain quiescence due to gas accretion from the inter-galactic medium, which
refuels star formation after $1-2\;\mathrm{Gyr}$. The 
major mergers examined in the four high resolution zoom in cosmological 
hydrodynamic simulation of \cite{Sparre:2016aa} also fail to sustain 
quiescence after $1-2\;\mathrm{Gyr}$ (Sparre et al. in prep.).
%Using the galaxy catalogs based on merger trees of the Millennium simulation (\citealt{De-Lucia:2007aa}),  \cite{Hirschmann:2013aa} find no significant difference between merger  histories of isolated galaxies and those in dense environments.  Although low and high density environments have significantly  different quiescent fractions, they do not have different merger histories.  Thus they conclude that mergers are not the main driver of quenching. 

AGN feedback has also been proposed as a quenching mechanism 
(\citealt{Kauffmann:2000aa, Croton:2006aa, Hopkins:2008ab,van-de-Voort:2011aa}), 
sometimes in conjunction with mergers as a way to sustain quiescence or on its
own. The feedback of the AGN deposit sufficient energy, which subsequently 
prevents additional gas from cooling. A number of more recent works have, 
however, cast doubt on the role of the AGN in quenching. \cite{Mendel:2013aa}, 
identified quenched galaxies, with selection criteria analogous to the 
selection of post-starburst galaxies, in the SDSS DR7 sample and found 
no excess of optical AGN in them, suggesting that AGN do not have 
defining role in quenching. \cite{Gabor:2014aa} further argue against AGN quenching 
by examining gas-rich, isolated disk galaxies in a suite of 
high resolution simulations where they find that the AGN outflows 
have little impact on the gas reservoir in the galaxy disk and 
furthermore fail to prevent gas inflow from the intergalactic medium.
\cite{Yesuf:2014aa} examined post-starburst galaxies transitioning from the 
blue cloud to the red sequence to find a significant time delay between the 
AGN activity and starburst phase, which suggests that AGN do not play a primary 
role in triggering quenching. AGN may yet be responsible for 
quenching in conjunction with other mechanisms or have a role 
in sustaining quiescence.

Besides mergers and AGN driven processes, another class of 
proposed mechanisms involves some process(es) that restrict the 
inflow of cold gas -- strangulation. With little
inflow of cold gas, the galaxy quenches as it depletes its 
cold gas reservoir. One mechanism that has been proposed 
to prevent cold gas accretion is loosely referred to as ``halo quenching''. 
A hot gaseous coronae, which form in halos with masses above 
$\sim 10^{12}\mathrm{M}_\odot$ via virial shocks, starves 
galaxies of cold gas for star formation (\citealt{Birnboim:2003aa, 
Keres:2005aa, Cattaneo:2006aa, Dekel:2006aa, Birnboim:2007aa, Gabor:2012aa, 
Gabor:2015aa}). 
For these sorts of mechanisms, the quenching timescale is 
linked to the time it takes for the galaxy to deplete its cold gas 
reservoir -- the gas depletion timescale. 

In principle, the gas depletion time can be estimated from 
measurements of the total gas mass or gas fraction. 
In \cite{Popping:2015aa}, for instance, they derive 
``star formation efficiency'' (SFE; inverse of the gas depletion time) 
by dividing the SFR of the SFMS by the total galaxy gas mass that 
they infer from their semi-empirical model. These sorts of gas 
depletion time estimates, however, have significant redshift 
dependence because the gas fraction of galaxies do not evolve 
significantly over $z < 1$ 
(\citealt{Stewart:2009aa, Santini:2014aa, Popping:2015aa}).
%\cite{Stewart:2009aa} and  \cite{Boselli:2014aa}, for instance, provide scaling relations  for $f_\mathrm{gas}$, which can be used to calculate the total gas mass from the galaxy stellar mass. The estimated gas  mass can be divided by the SFR of the SFMS for a gas depletion time  estimate. 

Nevertheless, in Figure~\ref{fig:tquench_comp} we estimate the  central quenching 
migration time ($t_\mathrm{mig}^\mathrm{cen}$; orange) -- the time it takes
central galaxies to migrate from the SFMS to quiescent estimated from 
our $\taucen$ -- to the gas depletion times derived from 
the \cite{Popping:2015aa} SFEs (blue). For $t_\mathrm{mig}^\mathrm{cen}$, 
we compute the time it takes a quenching galaxy to transition from 
the SFMS to the quiescent peak of the SFR distribution at $z = 0.2$. 
We compute $t_\mathrm{mig}^\mathrm{cen}$ at $z = 0.2$ because this 
is approximately when the $z \approx 0$ green valley galaxies would 
have started quenching. For the gas depletion time, we invert the 
SFE at $z = 0.2$, interpolated between the $z = 0.$ and $z=0.5$ 
\cite{Popping:2015aa} SFEs (blue dot-dashed). The surrounding 
blue shaded region marks the range of gas depletion times from 
$z = 0.15$ (longer) to $0.25$ (shorter) to illustrate the significant 
redshift dependence. We also note that over the redshift range 
$z = 0.5$ to $0.$, at $\mathcal{M} = 10^{10}\mathrm{M}_\odot$, 
the \cite{Popping:2015aa} gas depletion time varies from $\sim 2.5$ 
to $7 \;\mathrm{Gyrs}$. The $t_\mathrm{mig}^\mathrm{cen}$ and
gas depletion time in Figure \ref{fig:tquench_comp} are generally 
in agreement with each other and exhibit similar mass dependence.

Beyond the estimates of gas depletion times from gas mass, 
recently \cite{Peng:2015aa}, using a gas regulation model 
(e.g. \citealt{Lilly:2013aa, Peng:2014aa}), explored the impact that 
different quenching mechanisms have on the stellar metallicity 
of local galaxies from the SDSS DR7 sample. To reproduce the stellar
metallicity difference between quiescent and star forming galaxies 
in their galaxy sample, they conclude that the primary mechanism for 
quenching is gas depletion absent accretion and it a typical quenching 
migration time of $t_\mathrm{mig} \sim 4\;\mathrm{Gyr}$ for 
$\mathcal{M} < 10^{11}\mathrm{M}_\odot$. 
We infer the quenching migration time from Figure 2 of \cite{Peng:2015aa}
and include it in Figure \ref{fig:tquench_comp} (dashed). 
The \cite{Peng:2015aa} migration time exhibits a similar mass 
dependence as our central quenching migration time. Furthermore, 
although slightly shorter at $\mathcal{M} > 5 \times 10^{10} \mathrm{M}_\odot$, 
the migration time is broadly consistent with our central 
quenching migration time.

Overall, our $t_\mathrm{mig}^\mathrm{cen}$ is consistent with the 
migration time estimates of gas depletion mechanisms. In other words, 
our central galaxy quenching timescale is consistent with 
the timescales predicted by gas depletion absent accretion. One currently favored model 
for halting cold gas accretion -- halo quenching -- quenches galaxies 
that inhabit host halos with masses greater than some threshold 
$\sim 10^{12}\mathrm{M}_\odot$. Based on SHAM, this halo mass threshold 
corresponds to stellar masses of $\sim 10^{10.25} \mathrm{M}_\odot$. 
Yet, a significant fraction of the SDSS central galaxy population 
with stellar masses $< 10^{10.25} \mathrm{M}_\odot$ are quiescent.  
While, scatter in the halo mass threshold and the stellar mass to 
halo mass relation, combined, may help resolve this tension, halo quenching, 
faces a number of other challenges. For instance, the predictions of
halo quenching models are difficult to reconcile with the observed scatter
in the stellar mass to halo mass relation (\citealt{Tinker:2016aa}).
Furthermore, models that rely only on such ‘halo quenching’ still 
must account for the hot gas in the inner region of the halo, which, 
because of its high density, often has short cooling times of just $1-2~\mathrm{Gyr}$.
Of course, the challenges of halo quenching does {\em not} rule out 
quenching from gas depletion absent accretion since other mechanisms may 
also prevent cold gas from accreting onto the central galaxy 
%If halo quenching is the main driver of star formation quenching,  central galaxies in halos less massive than the threshold should rarely  be quenched. Based on SHAM, this halo mass threshold corresponds to  stellar masses of $\sim 10^{10.25} \mathrm{M}_\odot$; so central galaxies with $\mathcal{M}_* <  10^{10.25} \mathrm{M}_\odot$ should rarely be  quenched. In the SDSS DR7 group catalog, however, a significant fraction  of the central galaxy population with stellar masses  $< 10^{10.25} \mathrm{M}_\odot$ are quiescent.  

Finally, morphological quenching has also been proposed as a mechanism responsible 
for quenching star formation. In the mechanism proposed by \cite{Martig:2009aa}, 
for instance, star formation in galactic disks are 
quenched once the galactic disks become dominated by
a stellar bulge. This stabilizes the disk from fragmenting 
into bound, star forming clumps. In a cosmological zoom-in simulation of a 
$\sim 2 \times 10^{11} \mathrm{M}_\odot$ galaxy selected to examine such 
a mechanism, \cite{Martig:2009aa} finds that the galaxy quenches its star formation
from $\sim 10~\mathrm{M}_\odot\mathrm{yr}^{-1}$ to 
$\sim 1.5~\mathrm{M}_\odot\mathrm{yr}^{-1}$
in $\sim 2.5\;\mathrm{Gyr}$ during the morphological quenching phase. A 
$\mathcal{M} \sim 2 \times 10^{11} \mathrm{M}_\odot$ galaxy with 
$\hat{t}_\mathrm{Q} \sim 2.5\;\mathrm{Gyr}$ (star; Figure \ref{fig:tquench_comp}) 
is in good agreement with $\hat{t}_\mathrm{Q}^\mathrm{cen}$. 
Despite this agreement, morphological quenching faces a number 
of challenges. There is little evidence from modern cosmological
hydrodynamic simulations that suggest that morphological quenching can
drive anything beyond short timescale fluctuations in gas fueling and 
SFR. Furthermore, proposed morphological quenching mechanisms
face the ``cooling flow problem'' where they fail to prevent 
gas cooling onto a galaxy. Without addressing this issue, proposed 
morphological quenching mechanisms {\em cannot} maintain quiescence.

Our own Milky Way galaxy, as \cite{Haywood:2016aa} finds, after forming its
bar undergoes quenching. In the star formation history of the Milky Way 
that \cite{Haywood:2016aa} recovers, the SFR of the Milky Way decreases 
by an order of magnitude over the span of roughly 
$1.5\;\mathrm{Gyr}$. Converting to $\hat{t}_\mathrm{Q}$ in a similar fashion
as our $\hat{t}_\mathrm{Q}^\mathrm{cen}$ estimates and assuming a Milky Way stellar
mass of $\sim 6\times 10^{10}\mathrm{M}_\odot$ 
(\citealt{Licquia:2015aa, Haywood:2016aa}), 
we find remarkable agreement with our $\hat{t}_\mathrm{Q}^\mathrm{cen}$ 
(Figure \ref{fig:tquench_comp}). 
Motivated by the contemporaneous formation of the bar with quenching, 
\cite{Haywood:2016aa} suggest a bar driven (morphological) quenching mechanism 
that inhibits gas accretion through high level turbulence supported pressure 
that is generated from the shearing of the gaseous disk. Although, 
this proposal may resolve the cooling-flow problem, their arguments for the
mechanism are qualitative and thus require more detailed investigation. 
Admittedly, however, this particular comparison is hastily made since quenching 
event occurs beyond the redshift probed by our simulation at 
$1 < z < 2$. Furthermore, after dramatic quenching episode, based on 
the star formation history that \cite{Haywood:2016aa} recovers, the Milky Way
resumes star formation at a much lower level. 

The central quenching timescale we infer from our analysis provides key insight
into the physical processes responsible for quenching star formation. It offers 
a means of assessing the feasibility of numerous quenching mechanisms, which
operate on distinct timescale. Based on the latest models and simulations, 
merger driven quenching has fallen out of favor 
and AGN alone seem insufficient in triggering quenching. 
Mechanisms that halt cold gas accretion, such as halo quenching, predict 
quenching times generally consistent with our estimates from the central 
quenching timescale we derive. However, it fails to explain the significant 
low mass quiescent population of central galaxies. Morphological quenching, 
with its agreement in quenching time, may be a key physical mechanism 
in quenching star formation. However, more evidence is required that it 
can address the cooling flow problem and maintain quiescence. Furthermore, 
its role in the overall quenching of galaxy populations -- not just 
single simulated galaxies -- still remains to be explored.  

\section{Summary} \label{sec:summary}
Understanding the physical mechanisms responsible for quenching 
star formation in galaxies has been a long standing challenge 
for hierarchical galaxy formation models. Following the success 
of \cite{Wetzel:2013aa} in constraining the quenching timescales 
of satellite galaxies, in this work, we focus on star formation 
quenching in central galaxies with a similar approach. 
Using a high resolution $N$-body simulation in conjunction with 
observations of the SMF, SFMS, and quiescent fraction at $z < 1$, 
we construct a model that statistically tracks the star formation 
histories of central galaxies. The free parameters of our model 
dictate the height  of the green valley at the initial redshift, 
the correction to the quenching probability, and most importantly,
the quenching timescale of central galaxies, 

Using ABC-PMC with our model, we infer parameter constraints that best 
reproduce the observations of the central galaxy SSFR distribution from the 
SDSS DR7 Group Catalog and the central galaxy quiescent fraction evolution.
From the parameter constraints of our model, we find the following 
results: 
\begin{enumerate}
\item The quenching timescale of central galaxies exhibit a significant mass 
dependence: more massive central galaxies have shorter quenching timescales. 
Over the stellar mass range $\mathcal{M} = 10^{9.5} - 10^{11.5}\mathrm{M}_\odot$, 
$\taucen \sim 1.2 - 0.5 \;\mathrm{Gyr}$. Based on these timescales, 
central galaxies take roughly $2$ to $5$ Gyrs to traverse the green valley.  

\item The quenching timescale of central galaxies is significantly longer
than the quenching timescale of satellite galaxies. This result is robust for  
extreme prescriptions of the SMF evolution in our simulation and 
even for different parameterizations of the central quiescent fraction. 

\item The difference in quenching timescales of satellite and centrals
suggest that different physical mechanisms are primary drivers of 
star formation quenching in satellites versus centrals. Satellite galaxies  
experience external ``environment quenching'' while central galaxies  
experience internal ``self quenching''. 

\item We compare the central quenching timescales we infer to the 
gas depletion timescales predicted by quenching through strangulation 
and find broad agreement. We also find good agreement with 
morphological quenching; however, its feasibility in maintaining 
quiescent and for a wider galaxy population remains to be explored.

\end{enumerate}
\noindent Ultimately, the central galaxy quenching timescale we obtain in our analysis 
provides a crucial constraint for any proposed mechanism for star formation
quenching. 

One key component of our simulation is the use of SHAM to track evolution of
stellar masses of central galaxies. As mentioned above, the central galaxy 
quenching timescale results we obtain remain unchanged if we use stellar 
mass growth from integrated SFR. However, the use of SHAM stellar masses 
neglects the connection between stellar mass growth and star formation history. 
To incorporate integrated SFR galaxy stellar mass growth in our simulation, 
however, a better understanding of the detailed relationship among  
stellar mass growth, host halo growth, and the observed stellar mass to halo mass
relation is required. We will explore this in future work.  

\section*{Acknowledgements}
CHH was supported by NSF-AST-1109432 and NSF-AST-1211644. 
ARW was supported by a Moore Prize Fellowship through the 
Moore Center for Theoretical Cosmology and Physics at Caltech 
and by a Carnegie Fellowship in Theoretical Astrophysics at 
Carnegie Observatories. We thank Charlie Conroy and Michael R. Blanton for 
helpful discussions. CHH also thanks the Instituto  de F\'{i}sica 
Teo\'{o}rica (UAM/CSIC) and particularly Francisco Prada for 
their hospitality during his summer visit, where part of this 
work was completed.

\bibliographystyle{yahapj}
\bibliography{cenque}
\end{document}